\documentclass[aps,floats]{revtex4}
\usepackage{amsmath}
\usepackage{graphicx,epsfig}

\begin{document}
\bibliographystyle {plain}

\def\oppropto{\mathop{\propto}} 
\def\opsimeq{\mathop{\simeq}}
\def\opoverderline{\mathop{\overline}}
\def\operarrow{\mathop{\longrightarrow}}
\def\opsim{\mathop{\sim}}

\def\fig#1#2{\includegraphics[height=#1]{#2}}
\def\figx#1#2{\includegraphics[width=#1]{#2}}


\title{ Non-linear Response of the trap model in the aging regime : \\
Exact results in the strong disorder limit   } 

\author{C\'ecile Monthus}
\affiliation{Service de Physique Th\'eorique, 
Unit\'e de recherche associ\'ee au CNRS, \\
DSM/CEA Saclay, 91191 Gif-sur-Yvette, France}

\begin{abstract}

We study the dynamics of the one dimensional disordered trap model 
presenting a broad distribution of trapping times $p(\tau) \sim 1/\tau^{1+\mu}$,
when an external force is applied from the
very beginning at $t=0$, or only after a waiting time $t_w$,
in the linear as well as in the non-linear response regime.
Using a real-space renormalization procedure 
that becomes exact in the limit of strong disorder $\mu \to 0$,
we obtain explicit results for many observables,
such as the diffusion front, the mean position,
the thermal width, the localization parameters and
 the two-particle correlation function.
We discuss in details the various regimes 
that exist for the average position
in terms of the two times and the external field.

\end{abstract}

\maketitle

\section{Introduction}

 Trap models provide a simple phenomenological
mechanism for aging \cite{feigelman,jp92,reviewinapyoung}.
Their aging properties have thus been much studied,  
either in the mean field version 
\cite{jpdean,cmjp,barkai,bertinjptg}, where ``usual aging" occurs,
or in the one-dimensional version \cite{bertinjp,c_agingtrap}, where both
aging and subaging behaviors appear in different correlation functions.
The mathematicians have also been interested
by these trap models \cite{reviewbenarous}
with special attention for the cases 
$d=1$ \cite{isopi,benarous} and $d=2$ \cite{cerny}.
The one-dimensional version is moreover interesting on its own,
since it appears in various physical applications
concerning for instance transport properties in disordered chains
 \cite{alexander,reviewjpag}
or the dynamics of denaturation bubbles in random DNA sequences \cite{bubbledna}.

The study of the response to an external field and its relation with the
 thermal fluctuations has been for some years a central question
in the description of the aging dynamics of glassy systems 
\cite{reviewleticia,reviewritort}.
It is thus natural to consider the trap models from this point of view.
The studies on the violation of the
fluctuation-dissipation relation in mean-field trap models have shown that
the results depend on the observable \cite{fieldingsollich},
and on the choice of functional form of the hopping
rates \cite{ritort,sollich}.
For trap models on an hypercubic lattice, there are no such ambiguities
in the choice of observables and external fields,
since the natural observable is the position : one is 
interested into the response of 
 the position to an external bias,
and in the thermal fluctuations of the position. 
Recently, the response of the one-dimensional trap model was
studied via scaling arguments and numerical
simulations in \cite{bertinjpreponse}, where various regimes were found
depending on the relative values of the two times considered $(t_w,t_w+t)$
and the external applied field $f$, the main results being that 
in the linear response regime,
the Fluctuation-Dissipation Relation (or Einstein's relation)
is still valid in the aging sector,
whereas the response always become non-linear at long times.
We have shown in \cite{c_nonlinearft}
that these two response properties 
could be understood as consequences of a ``non-linear Fluctuation Theorem"
that has for origin a very special dynamical property
of the trap model.

In this paper, we consider again the response properties 
of the trap model, but with a complementary point of view :
we use a real-space renormalization group
(RSRG) procedure to derive various explicit exact results
in the limit of high disorder. 
The RSRG methods, that have appeared in the field of
disordered quantum spin chains
\cite{madasgupta,daniel}, have then been very
powerfull to study the Sinai model
\cite{us_sinai}, as well as reaction-diffusion processes
in a Brownian potential \cite{us_reactiondiffusion}.
In particular, since the out-of-equilibrium
dynamics of the random field Ising model (RFIM)
can be describded as a reaction-diffusion process 
in a Brownian potential for the domain walls,
the RSRG method has been used \cite{us_rfim} to study the response
of the RFIM to an applied external magnetic field. 
The RSRG approach is also very appropriate to study one-dimensional trap models
with a broad distribution of trapping times $p(\tau) \sim 1/\tau^{1+\mu}$
in the limit of high disorder $\mu \to 0$, as explained in details in
\cite{c_directed} for the {\it directed} trap model,
and in \cite{c_agingtrap} for the {\it symmetric} trap model
(i.e. in the absence of an external bias ) :
the RSRG method is able to reproduce
 the exact exponents of the whole aging phase
$0<\mu<1$ and moreover allows to compute exact scaling functions
for all observables in a systematic perturbation expansion
in $\mu$ \cite{c_directed,c_agingtrap}.
In contrast with other usual methods for disordered systems, 
the disorder average is not performed at the beginning
but at the very end : the RSRG procedure is defined 
sample by sample, all observables are then evaluated
in terms of the relevant properties of a given sample,
 and can be then averaged 
with the appropriate measure over the samples. 
The RSRG approach thus provides a very clear insight
into the important dynamical processes.

In this paper, we generalize the RSRG approach
describded in \cite{c_agingtrap} for the {\it unbiased} trap model
to include the influence of an external bias,
and we obtain exact results for various observables
in the high disorder limit $\mu \to 0$.

The paper is organized as follows.
In Section \ref{models}, 
we describe the trap model in an external force field $f$.
In Section \ref{rgprocedure}, we explain
 the real space renormalization procedure
in the presence of an external field. In Section \ref{dynamicsft}, we 
describe the effective dynamics when the external field is applied
from the initial time $t=0$, and we give explicit results
for 
one-time observables in Section \ref{observablest0}.
In Section \ref{dynamicsfttw}, we discuss the effective dynamics
when the external field is applied
only after a waiting time $t_w$, and we compute
the corresponding two-time observables in Section \ref{observablesttw}.
In Section \ref{reponseeffective}, we discuss the various regimes existing for the disordered averaged mean position in terms of the two times
 and the external field. In Section \ref{rareevents},
we discuss the rare events that are responsible for the response,
in the time sector where the effective dynamics gives no contribution.
Finally, in Section \ref{fdtsection}, we compare the response and
the thermal fluctuations in a given sample,
to discuss the validity of the
Fluctuation-Dissipation Relation.
The conclusions are given in
Section \ref{conclusion},
and the Appendices contain more technical details.

\section{ Models and notations }

\label{models}

\subsection{ Master Equation in a sample}

To study the trap model in an external force field $f$,
we conside the following Master Equation \cite{reviewjpag,c_nonlinearft}
\begin{eqnarray}
\frac{dP_t^{(f)}(x)}{dt} =   P_t^{(f)}(x+1) W_{ \{x+1 \to x \}}^{(f)}
+ P_t^{(f)}(x-1) W_{ \{x-1 \to x \}}^{(f)} 
-    P_t^{(f)}(x) \left[ W_{ \{x \to x+1 \}}^{(f)}
 + W_{\{ x \to x-1 \}}^{(f)} \right]
\label{master}
\end{eqnarray}
with the initial condition
$P_{t=0}^{(f)}(x)=\delta_{x,0}$.
The hopping rates 
\begin{eqnarray}
W_{ \{x \to x \pm 1 \}}^{(f)} && = e^{- \beta E_{x} \pm \beta \frac{f}{2} }
\end{eqnarray}
 satisfy the detailed balance condition
\begin{eqnarray}
\frac{W_{ \{x \to x+1 \}}}{W_{ \{x+1 \to x \}}}
= e^{ \beta( U_x-U_{x+1} )}
\label{detailedbalance}
\end{eqnarray} 
where the total energy 
\begin{eqnarray}
 U_x = -E_x -f x 
\label{defui}  
\end{eqnarray}
 contains both the random energy $(-E_x)$ of the trap $x$
and the potential energy $(-f x)$ linear in the position $x$
induced by the external applied field $f$.

\subsection{Law for the disorder}

The trap energies $\{E_x\}$ are quenched random variables
distributed exponentially \cite{jp92}
\begin{eqnarray}
\rho(E)= \theta(E>0) \frac{1}{T_g} e^{- \frac{E}{T_g}} 
\label{rhoe} 
\end{eqnarray}
This corresponds
for the mean trapping time $\tau = e^{\beta E}$
to the algebraic law
\begin{eqnarray}
q(\tau)
= \theta(\tau>1)  \frac{\mu}{\tau^{1+\mu}} 
 \label{qtau}
\end{eqnarray}
with the temperature-dependent exponent 
$\mu = \frac{T}{T_g}$
At low temperatures $\mu<1$, the mean trapping time $\int d \tau 
\tau q(\tau)$ is infinite 
and this directly leads to aging effects. 

\subsection{Link with the `trap model with asymetry'}

To make the link with the `trap model with asymetry h' studied in \cite{bertinjpreponse}, 
we note that in the new time ${\tilde t} = ( 2 \cosh \beta \frac{f}{2} ) t $,
the Master equation (\ref{master}) becomes
\begin{eqnarray}
 \frac{dP_{\tilde t}^{(f)}(x)}{d{\tilde t}} =  \frac{ q_-(f)}{\tau_{n+1} } P_{\tilde t}^{(f)}(x+1)
+ \frac{ q_+(f)}{\tau_{n-1} } P_{\tilde t}^{(f)}(x-1)  
-  \frac{ 1 }{\tau_n}  P_{\tilde t}^{(f)}(x) 
\label{master2}
\end{eqnarray}
where 
\begin{eqnarray}
q_{\pm}(f)  = \frac{ e^{ \pm \beta \frac{f}{2} }}
{e^{ + \beta \frac{f}{2} }+e^{ - \beta \frac{f}{2} }} = \frac{1\pm h(f)}{2} 
\label{q+-f}
\end{eqnarray}
are the probabilities to jump on the right and on the left when escaping
a trap, with the normalization $q_+ + q_-=1$. The asymetry 
\begin{eqnarray}
h(f)=q_+-q_- = \tanh  \beta \frac{f}{2}  
\end{eqnarray}
varries between $h(f=0)=0$ for the unbiased case and $h(f \to \infty) \to 1$
for the fully directed case.

As soon as $h>0$, the random walk is expected to become
asymtotically directed on large scales. 
 In this article, we will be interested into the case 
where the crossover towards the directed regime 
happens on large length scales, i.e. the 
local asymetry is very small $h \ll 1$
or equivalently $\beta f \ll 1$. In this regime also considered
in \cite{bertinjpreponse},
the relation between the force and the asymetry
is at lowest order simply linear 
\begin{eqnarray}
h(f)=\frac{\beta f}{2 } + O \left( ( \beta f )^3 \right) 
\label{hof} 
\end{eqnarray}
and thus the results of the present paper can be straighfordwardly
compared with \cite{bertinjpreponse}.

\subsection{ Entropy and Generalized free-energy  }

As in similar models \cite{kurchan,bounds,us_energysinai},
the Shannon entropy 
\begin{eqnarray}
S(t) = - \sum_{x} P_{t}(x) \ln P_{t}(x)
\label{defentropy}  
\end{eqnarray}
and the energy $ U(t) =  \sum_{x} P_{t}(x) U_x $ (\ref{defui})  
allow to define a generalized free-energy 
\begin{eqnarray}
F(t) = U(t) -T S(t) = \sum_{x} P_{t}(x) \left[ U_x + T \ln P_{t}(x) \right]  
\end{eqnarray}
The detailed balance condition implies
that it is a non-increasing function
 $\frac{ dF(t) }{dt}  \leq 0 $
The equality with zero is possible only if 
all currents exactly vanish, corresponding to equilibrium. 
Here, since we consider the infinite line,
the equilibrium cannot be reached and the free-energy will decrease with no bounds.

\section{ Real space renormalization procedure in the presence of a field}

\label{rgprocedure}

We have already presented the real space renormalization procedure  
 for the {\it unbiased } trap model in \cite{c_agingtrap}. 
 Here we wish to generalize this approach to the presence of
an external bias $f>0$.

\label{rglandscape}

\subsection{Notion of renormalized landscape at a scale $R$}

The basic idea of the Real Space Renormalization procedure
\cite{us_sinai,c_directed,c_agingtrap}
is that the dynamics at large time is dominated by the
statistical properties of the large trapping times.
The renormalized landscape at scale $R$
is defined as follows : all traps with trapping time $\tau_n<R$
are decimated and replaced by a ``flat landscape'', whereas 
all traps with waiting time $\tau_n>R$ remain unchanged.
At large scale $R$, the distribution of the distance $l$ between two traps
of the renormalized landscape at scale $R$ 
takes the scaling form  
\begin{eqnarray}
P_R(l) \simeq \frac{1}{R^{\mu} } {\cal P} 
\left(\lambda= \frac{l }{R^{\mu} } \right)
\label{length}
\end{eqnarray}
where the scaling distribution is simply exponential
${\cal P} (\lambda)= e^{- \lambda} $

The distribution of the trapping times of 
the traps in the renormalized landscape at scale $R$ 
is simply
\begin{eqnarray}
q_R(\tau)
 = \theta(\tau>R) \frac{\mu}{\tau} \left( \frac{R}{\tau} \right)^{\mu}
 \label{qrtau}
\end{eqnarray}
To relate the renormalization scale $R$ to the time $t$, we 
have to study the time needed to escape 
from a renormalized trap. 

\subsection{ `Escape time' from a renormalized trap 
to another renormalized trap }

\begin{figure}

\centerline{\includegraphics[height=8cm]{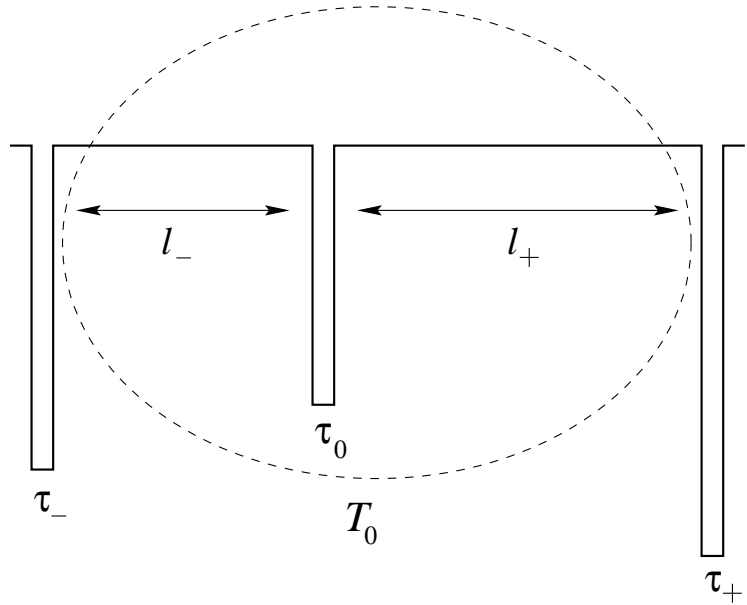}} 
\caption{ Definition of the escape time from a trap in the renormalized 
landscape :
the trap of escape time $\tau_0$ existing 
in the renormalized landscape at scale $R$
is surrounded by two
renormalized traps that are at distances $l_+$
and $l_-$ on each side. The escape time $T_0$ is
the mean time needed to reach either $\tau_+$
or $\tau_-$ when starting at $\tau_0$.   } 
\label{defescape}
\end{figure}

We now study the `escape time' $T$
from a trap $\tau_0$ existing in the renormalized landscape at scale $R$
in the presence of a field $f>0$. 
This trap is surrounded by two
renormalized traps that are at distances $l_+$
and $l_-$ on each side (see Figure \ref{defescape}).
Whenever the particle escapes from the trap $\tau_0$, 
it escapes on the right with probability $q_+=(1+h(f))/2$
and on the right with probability $q_-=(1-h(f))/2$ (\ref{q+-f}).

If it escapes on the right,
it will succeed to reach the trap $\tau_+$
with probability (\ref{escape+}) 
\begin{eqnarray}
p_e^+(f,l_+)=   \frac{  1 - e^{- \beta f  }  }
{1- e^{- \beta f l_+ } } 
\label{escape+t}
\end{eqnarray}
which varries between $p_e^+(l,f \to 0)=1/l_+$ for the unbiased case
and $p_e^+(l,f \to \infty)=1$ for the directed case.

If it escapes on the left, it will succeed to reach the trap $\tau_-$
with probability (\ref{escape-}) 
\begin{eqnarray}
p_e^-(f,l_-) =  \frac{ ( 1 - e^{- \beta f  } ) e^{- \beta f (l_- -1) }  }
{1- e^{- \beta f l_- } } 
 \label{escape-t}
\end{eqnarray}
which varries between $p_e^-(l_-,f \to 0)=1/l_-$ for the unbiased case
and $p_e^-(l_-,f \to \infty)=0$ for the directed case.

 Otherwise, it will be 
re-absorbed again by the trap $\tau_0$.
So the total probability to escape when exiting from $\tau_0$ reads
\begin{eqnarray}
p_e(f,l_+,l_-) = \frac{ e^{ + \beta \frac{f}{2} } p_e^+(l_+,f)
 + e^{ - \beta \frac{f}{2} }  p_e^-(l_-,f)}
{e^{ + \beta \frac{f}{2} }+e^{ - \beta \frac{f}{2} }} 
\label{pescape}
\end{eqnarray}

 \subsubsection{ Number of sojourns in a renormalized
 trap before escape to a neighbor renormalized trap}

 As a consequence, the probability $E_R(n)$ to 
need $(1+n)$ successive sojourns in the trap $\tau_0$ before the particle
succeeds to escape either to the trap $\tau_-$ or to the trap $\tau_+$ reads   
\begin{eqnarray}
E_R(n) && = \left[1- p_e(f,l_+,l_-) \right]^{n}
 p_e(f,l_+,l_-)
\label{distrin}
\end{eqnarray}

For large $R$, since we have
$l_{\pm}=R^{\mu} \lambda_{\pm}$  (\ref{length})
we obtain using (\ref{scalingescape+},\ref{scalingescape-})
the scaling form for the probability of escape (\ref{pescape})
\begin{eqnarray}
p_e(f,l_+,l_-) \opsimeq \frac{1}{R^{\mu}} \phi(F\equiv \beta f R^{\mu}
; \lambda^+ = \frac{l_+}{R^{\mu}}
,\lambda^-= \frac{l_-}{R^{\mu}})
\end{eqnarray}
with the scaling function
\begin{eqnarray}
 \phi(F; \lambda^+,\lambda^-)
=   \frac{ F (1-e^{- F ( \lambda^++\lambda^-) })}{ 2 (1-e^{- F  \lambda^+ }) (1-e^{- F \lambda^- }) }
\label{defphi}
\end{eqnarray}
The expansion near $F \to 0$
gives the first correction with respect to the symmetric case
studied in \cite{c_agingtrap}
\begin{eqnarray}
 \phi(F; \lambda^+,\lambda^-) =
\frac{1}{2} \left[   \frac{ 1 }{   \lambda^+ }
+ \frac{ 1 }{  \lambda^- } \right]
+ \frac{ \lambda^+ + \lambda^-}{24} F^2
-  \frac{ (\lambda^+)^3 + (\lambda^-)^3}{1440} F^4+O(F^6)
\end{eqnarray}
whereas in the other limit $F \to \infty$, we have 
\begin{eqnarray}
 \phi(F; \lambda^+,\lambda^-) \opsimeq_{F \to \infty} 
  \frac{F}{2} \left( 1+ e^{- F  \lambda^+ } + e^{- F \lambda^- }+...  \right) 
\end{eqnarray}

As a consequence, at large scale $R$, the number $n$ of returns
is distributed exponentially
\begin{eqnarray}
E_R(n) \simeq  \frac{1}{ <n>(R,F)  } e^{- \frac{n}{ <n>(R,F)  } } 
\label{lawn}
\end{eqnarray}
where the mean number of returns
\begin{eqnarray}
<n>(R,F) = \frac{ R^{\mu} }{ \phi(F; \lambda^+,\lambda^-) } 
\end{eqnarray}
varries between
\begin{eqnarray}
<n>(R,F \to 0) = R^{\mu} 
\frac{ 2 \lambda^+ \lambda^-}{   \lambda^+ + \lambda^-  } 
\end{eqnarray}
and
\begin{eqnarray}
<n>(R,F \to \infty) \simeq  R^{\mu} 
\frac{ 2 }{ F } =   
\frac{ 2 }{ \beta f }
\end{eqnarray}

 \subsubsection{ Total time spent inside a renormalized trap before escape to a neighbor renormalized trap}

Let us now consider the probability distribution $P_{in}(t_{in})$
of the total time $t_{in}$ spent inside the trap $\tau_0$ before its escape.
It can be decomposed into the number $n$ of sojourns,
where $n$ is distributed with (\ref{distrin})
\begin{eqnarray}
t_{in} = \sum_{n=1}^{1+n} t_n 
\end{eqnarray}
where $t_n$ is the time spent during the sojourn $i$
in the trap $\tau_0$, so it is distributed with
the exponential distribution with mean time $\tau_0$.
Actually, since $n$ is large in the large $R$ limit,
we have the central-limit theorem
\begin{eqnarray}
t_{in} \opsimeq_{n \to \infty} n <t_n> = n \tau_0
\label{tin}
\end{eqnarray}
Since the number $n$ is distributed with
exponentially (\ref{lawn}),
we finally obtain that $t_{in}$ is also exponentially distributed 
\begin{eqnarray}
{\hat P}_{in}(t_{in}) 
\opsimeq_{R \to \infty}  \frac{1}{T_0} e^{- \displaystyle \frac{t_{in}}{T_0} }
\label{pin}
\end{eqnarray}
with the characteristic time
\begin{eqnarray}
T_0 = \tau_0 <n>(R,F) 
\end{eqnarray}

Since the smallest trapping times 
existing in the renormalized landscape at scale $R$ is $\tau_0=R$,
the time spent inside the trap $\tau_0$ before it succeeds
to escape scales as
 \begin{eqnarray}
t_{in}(R,F) \opsim_{R \to \infty}  R <n>(R,F) = \frac{ R^{1+\mu} }{ \phi(F; \lambda^+,\lambda^-) }  
\label{scaletin}
\end{eqnarray}

 \subsubsection{ Total time spent during the unsuccessful excursions before the escape}

Among the $n$ unsuccessful
excursions, there are $m$ excursions on the left
and $(n-m)$ excursions on the left, where $m$ is distributed with 
the binomial distribution $2^{-n} C_n^m$. 
Since $n$ and $m$ are large, we again have a central-limit theorem
 \begin{eqnarray}
t_{out} = \sum_{n=1}^m t_n^- +\sum_{j=1}^{n-m} t_j^+ \opsimeq
m \theta_{us}^-(f,l_-) + (n-m) \theta_{us}^+(f,l_+) 
\label{tout} 
\end{eqnarray}
where $\theta_{us}^{\pm}(f,l)$ 
represents the mean time needed to return to $0$ when starting
at $(\pm 1)$ without touching the point $(\pm l)$ 
in a flat landscape. Using the asymptotic behavior 
(\ref{meanplus},\ref{samepm})
\begin{eqnarray}
\theta_{us}^+(f,l) \opsimeq_{l \to \infty} l \Theta(u=\beta fl)
\end{eqnarray}
with the scaling function $\Theta$ given in (\ref{thetaexcursion}),
the ratio between $t_{out}$ 
and $t_{in}$ (\ref{tin}) scales as
 \begin{eqnarray}
\frac{ t_{out} }{t_{in}}(R,F)
\simeq \frac{ \theta_{us}^-(f,l_-) +  \theta_{us}^+(f,l_+)}{ 2 \tau_0 } 
\sim R^{\mu-1} \left[ \Theta( F  \lambda^+) + \Theta( F \lambda^-) \right]     
\label{scaletout}
\end{eqnarray}
which varries between (\ref{theta0})
 \begin{eqnarray}
\frac{ t_{out} }{t_{in}}(R,F \to 0)
\sim  R^{\mu-1} 
\end{eqnarray}
and 
 \begin{eqnarray}
\frac{ t_{out} }{t_{in}}(R,F \to \infty)
\sim R^{\mu-1} \frac{1}{F} = \frac{1}{R \beta f}
\end{eqnarray}
We may thus neglect $t_{out}$ with respect to $t_{in}$ at large scale $R$.

 \subsubsection{ Time spent during the successful excursion to escape}

We finally consider the diffusion time $t_{diff}^{\pm}$
 of the successful escape to the neighbor renormalized landscape
$(\pm l)$ when starting at $x=1$
without visiting $x=0$. 
The mean time needed to reach $x=l_+$
when starting at $x=1$ for a random walk conditioned not to visit $x=0$
takes the scaling form (\ref{diffusionlf}
\begin{eqnarray}
t_{diff}^+(f,l_+) \simeq l_+^2 D( u=\beta fl_+) ) 
\label{tdifffl}
\end{eqnarray}
where the scaling function $D(u)$ is given in (\ref{functionDu}).
Using also (\ref{diff-+}), we obtain the scaling
 \begin{eqnarray}
\frac{ t_{diff} }{t_{in}}(R,F)
\sim  R^{ \mu-1 } {\cal D} ( F \lambda ) 
 \label{scaletdiff}
\end{eqnarray}
which is the same as (\ref{scaletout})
and thus $t_{diff}$ is again negligible with respect to $t_{in}$ (\ref{scaletin}).

\subsubsection{ Conclusion}

So we obtain that the total time 
\begin{eqnarray}
t_{esc} = t_{in}+t_{out}+t_{diff}
\label{tesc3contri}
\end{eqnarray}
 needed to escape is actually simply given by
the time $t_{in}$ spent inside the trap $\tau_0$.
So the distribution of $t_{esc}$
is given by the exponential (\ref{pin})
with the escape time $T_0$.

In conclusion, a trap 
 of the renormalized landscape at scale $R$
has a trapping time $\tau$ distributed with (\ref{qrtau}), but has 
an `escape time' proportional to $\tau$ 
\begin{eqnarray}
T= \tau <n>(R,F) = \tau \frac{ R^{\mu} }{ \phi(F; \lambda^+,\lambda^-) } 
\end{eqnarray}
in terms of the two rescaled distances $\lambda_{\pm}$ to the neighbors.

\subsection{Distribution of `escape times' in the renormalized landscape}

The distribution of the escape time $T$ in the renormalized landscape
at scale $R$ is obtained by averaging over $\tau,\lambda_{\pm}$
\begin{eqnarray}
Q_R(T;f) && = \int_R^{+\infty} d\tau \frac{\mu}{\tau} \left( \frac{R}{\tau} \right)^{\mu} \int_0^{+\infty} d  \lambda^+ d \lambda^- e^{-  \lambda^+-\lambda^-}
\delta(T- \frac{R^{\mu}}{\phi(F; \lambda^+,\lambda^-)}  \tau) 
\end{eqnarray}
It thus takes the scaling form
\begin{eqnarray}
Q_{R}(T;f) && = \frac{1}{R^{1+\mu}} {\cal Q}_{\mu} \left( {\tilde T} = \frac{T}{R^{1+\mu}}; F \right)
\label{qscaling}
\end{eqnarray}
where the scaling function
\begin{eqnarray}
{\cal Q}_{\mu} \left( {\tilde T};F \right)
&& = \frac{\mu}{(\tilde T)^{1+\mu}} 
\int_0^{+\infty} d  \lambda^+ d \lambda^- e^{-  \lambda^+-\lambda^-}
\left[ \phi(F; \lambda^+,\lambda^-) \right]^{-\mu}
\theta\left[ \phi(F; \lambda^+,\lambda^-) > \frac{1}{\tilde T} \right]
\end{eqnarray}
presents the asymptotic behavior 
\begin{eqnarray}
{\cal Q}_{\mu} \left( {\tilde T};F \right)
 && \opsimeq_{{\tilde T} \to \infty}
 \frac{\mu}{(\tilde T)^{1+\mu}} c_{\mu}(F)
\label{qrtasymp}
\end{eqnarray}
with the prefactor
\begin{eqnarray}
 c_{\mu}(F) = \int_0^{+\infty} d  \lambda^+ d \lambda^- e^{-  \lambda^+-\lambda^-}
\left[ \phi(F; \lambda^+,\lambda^-) \right]^{-\mu}
= 1 - \mu \int_0^{+\infty} d  \lambda^+ d \lambda^- e^{-  \lambda^+-\lambda^-}
\ln \left[ \phi(F; \lambda^+,\lambda^-) \right] +O(\mu^2)
\label{calcmu}
\end{eqnarray}

For small $\mu$, 
the probability distribution ${\cal Q}_{\mu} \left( {\tilde T},F \right)$ is dominated by its long tail, and we may approximate
it by
\begin{eqnarray}
{\cal Q}_{\mu} \left( {\tilde T} \right) 
\simeq \theta( {\tilde T} > {\tilde T}_{\mu} (F) )
\frac{\mu}{{\tilde T}} \left( \frac{{\tilde T}_{\mu} (F)}
{{\tilde T}}\right)^{\mu} 
\label{qpower}
\end{eqnarray}
The cut-off ${\tilde T}_{\mu}(F)$ chosen
 to preserve the normalization
is determined by the coefficient in
the long tail part (\ref{qrtasymp},\ref{calcmu}) 
\begin{eqnarray}
 {\tilde T}_{\mu}(F) = \left( c_{\mu}(F)  \right)^{\frac{1}{\mu} } 
= e^{S(F)}+ O(\mu)
\end{eqnarray}
where the function $S(F)$ obtained from (\ref{calcmu})
can be computed with the explicit expression (\ref{defphi})
\begin{eqnarray}
S(F) && \equiv -  \int_0^{+\infty} d  \lambda^+ d \lambda^- e^{-  \lambda^+-\lambda^-}
\ln \left[ \phi(F; \lambda^+,\lambda^-) \right] \\
&& = - \ln F + \ln 2 - \gamma_E- \psi \left( 1+ \frac{1}{F} \right)
- \frac{1}{F} \psi' \left( 1+ \frac{1}{F} \right) 
\label{s0h}
\end{eqnarray}
in terms of $\psi(z)=\Gamma'(z)/\Gamma(z)$ and the Euler constant 
$\gamma_E=-\psi'(1)$.

For the unrescaled probability distribution (\ref{qscaling}),
this corresponds to the cut-off
\begin{eqnarray}
T_{0}(R,f)= R^{1+\mu} e^{ S( F ) }
\label{torh}
\end{eqnarray}

\subsection{Choice of the renormalization scale $R$ as a function of time}

The renormalization scale $R$ has to be choosen as a function of time by the requirement that this effective cut-off $T_0(R,f)$
is exactly $t$, 
meaning that at time $t$, only traps with escape times $T>t$
have been kept, whereas all traps with escape times $T<t$
have been removed and replaced by a flat landscape.
So given a renormalisation scale $R$ and an external field $f$,
the corresponding time $t$ of the dynamics reads
\begin{eqnarray}
t=t(R,f) \equiv R^{1+\mu} e^{ S(F= \beta f R^{\mu} ) }
\end{eqnarray}
The renormalization scale $R(t,f)$ as a function of time $t$
and field $f$ is thus defined by the implicit equation
\begin{eqnarray}
t = R^{1+\mu}(t,f) e^{ S( \beta f R^{\mu}(t,f) ) }
\label{choicet}
\end{eqnarray}

At short times, assuming $F= \beta f R^{\mu}(t) \ll1$, we may use the expansion
\begin{eqnarray}
S(F)= \ln 2 -1  - \gamma_E - \frac{1}{12} F^2 +O(F^4)
\end{eqnarray}
to solve (\ref{choicet}) to
obtain the first correction with respect to the unbiased case
\begin{eqnarray}
R(t,f) = \left(  \frac{t}{T_0} \right)^{\frac{1}{1+\mu}}
\left[ 1 + \frac{(\beta f)^2}{12 (1+\mu)} \left(  \frac{t}{T_0} \right)^{\frac{2 \mu }{1+\mu}}  ... \right] 
\label{roftfsmall}
\end{eqnarray}
where ${\tilde T_0}=e^{S(0)}=2 e^{-1-\gamma_E}$. 
This solution is thus valid for  
\begin{eqnarray}
t \ll t_{\mu}(f) \equiv \left( \frac{1}{ \beta f } \right)^{\frac{1+\mu}{\mu} }
\label{tmuf}
\end{eqnarray}
where the time scale $t_{\mu}(f)$ was already  
shown to play an important role 
via other approaches \cite{bertinjpreponse,c_nonlinearft}.

At long times, assuming $F=\beta f R^{\mu}(t) \gg1$, we may use the 
asymptotic behavior
\begin{eqnarray}
S(F) = - \ln \frac{F}{2} - \frac{\pi^2}{3  F}
 +O(\frac{1}{F^2})
\end{eqnarray}
to solve (\ref{choicet}) which leads to
\begin{eqnarray}
R(t,f) = \frac{\beta f}{2} t \left[ 1 +\frac{\pi^2 2^{\mu}}
{3 (\beta f)^{1+\mu} t^{\mu} } 
+ ...  \right]
\end{eqnarray}
which is valid for $t \gg t_{\mu}(f)$ (\ref{tmuf}).

As a consequence, the characteristic length scale
corresponding to the mean distance between renormalized traps at scale $R(t)$
(\ref{length}) 
\begin{eqnarray}
\xi(t,f) = R^{\mu}(t,f)
\label{xit}
\end{eqnarray}
behaves as in the unbiased case at short times  
\begin{eqnarray}
\xi(t,f) \opsimeq_{t \ll t_{\mu}(f)} t^{\frac{\mu}{1+\mu}}
\left[ 1+O(\mu)  \right]
\label{xifsmall} 
\end{eqnarray}
and as  
\begin{eqnarray}
\xi(t,f) \opsimeq_{t \gg t_{\mu}(f)} (\beta f t)^{\mu} \left[  1+O(\mu) \right]
\label{xifbig}
\end{eqnarray}
at long times.

\section{Effective dynamics in the limit $\mu \to 0$}

\label{dynamicsft}

\subsection{ Probabilities to escape on the right or on the left
in the renormalized landscape }

We are now interested into the relative probability to escape on the right 
rather than to the left which reads (\ref{escape+t},\ref{pescape})
using (\ref{scalingescape+},\ref{scalingescape-})
\begin{eqnarray}
w_+(f,l_+,l_-) \equiv \frac{ e^{ + \beta \frac{f}{2} } p_e^+(f,l_+)}{ p_e(f,l_+,l_-) }
\opsimeq W_+(F,  \lambda^+, \lambda^-)
\end{eqnarray}
with the scaling function
\begin{eqnarray}
 W_+(F,  \lambda^+, \lambda^-) = \frac{1-e^{-F \lambda^-} }
{ 1- e^{-F ( \lambda^++\lambda^-) }}
\label{formw+}
\end{eqnarray}
The complementary probability to escape on the left reads
\begin{eqnarray}
 W_-(F,  \lambda^+, \lambda^-) = \frac{ e^{-F \lambda^-}(1-e^{-F  \lambda^+}) }
{ 1- e^{-F ( \lambda^++\lambda^-) }}
\label{formw-}
\end{eqnarray}

The expansion near $F \to 0$
gives the first correction with respect to the symmetric case
\cite{c_agingtrap}
\begin{eqnarray}
 W_+(F,  \lambda^+, \lambda^-) && = \frac{\lambda^-}
{ \lambda^- +  \lambda^+}
+ \left[   \frac{ 1 }{   \lambda^+ }
+ \frac{ 1 }{  \lambda^- } \right] \frac{F}{2}
+ \frac{ \lambda^+ - \lambda^-} {12} \left[   \frac{ 1 }{   \lambda^+ }
+ \frac{ 1 }{  \lambda^- } \right] F^2+O(F^3)
\label{w+fsmall}
\end{eqnarray}
whereas for large $F$, the first corrections with respect to the directed case
are given by
\begin{eqnarray}
 W_+(F,  \lambda^+, \lambda^-) \opsimeq_{F \to \infty}
1- e^{- F \lambda^-} + e^{- F ( \lambda^++\lambda^-) }+...
\end{eqnarray}

\subsection{ Rules for the effective dynamics}

We thus define the effective dynamics by the following rules :

The particle starting at the origin $O$ 
 will be at time $t$ either in the first trap $M_+$ of the renormalized landscape at scale $R(t,f)$
 on its right or in the 
 first trap $M_-$ of the renormalized landscape
 on its left.
 The weights of the traps $M_+$ and $M_-$ are given in terms of (\ref{formw+},\ref{formw-}) 
 by 
 \begin{eqnarray}
p_{[M_-M_+]}(M_+ \vert 0) && = W_+(F,  \lambda^+, \lambda^-)  \\
p_{[M_-M_+]}(M_- \vert 0) && = W_-(F,  \lambda^+, \lambda^-) 
\label{rulespplusmoins}
\end{eqnarray}
We now verify that this effective dynamics
presents some important properties.

\subsection{ Consistency upon iteration}

The rule for the effective dynamics is consistent upon iteration.
Indeed, suppose there are three consecutive traps  :
the trap $M_-$ is at a distance $l_-$ from the origin on the left,
the trap $M_+$ is at a distance $l_+$ from the origin on the right,
and the trap $M_{++}$ is at a distance $l$ from the trap $M_+$ on the right.

Suppose that the trap $M_+$ is decimated before the traps $M_-$ and $M_{++}$.
The new weights for the traps $M_-$ and $M_{++}$ become
 \begin{eqnarray}
p_{M_-}' && = p_{[M_-M_+]}(M_- \vert 0) + 
p_{[M_-M_{++}]} (M_- \vert M_+) p_{[M_-M_+]}(M_+ \vert 0) 
 = p_{[M_-M_{++}]}(M_- \vert 0) \\
p_{M_{++}}' && =  p_{[M_-M_{++}]}(M_{++} \vert M_+) 
p_{[M_-M_+]}(M_+ \vert 0) 
= \frac{ 1-e^{- F l_-} }{ 1- e^{- F (l_++l_-+l) } }  = p_{[M_-M_{++}]}(M_{++} \vert 0)
\label{ruledecim}
\end{eqnarray}
and thus the rules (\ref{rulespplusmoins}) 
for the occupancies of renormalized traps
are consistent upon decimation of traps in the renormalized landscape.

\subsection{ Conservation of the mean position in any sample for $f=0$}

As already emphasized in \cite{bertinjpreponse,c_nonlinearft}, the trap model
has a very special property : when there is no external field $f=0$
the mean position is a conserved quantity in any given sample,
as a simple consequence of the master equation (\ref{master})
for any realization of the trapping times $\{\tau_n\}$.
Here in the effective dynamics, the mean position
vanishes indeed in each sample with the 
weights in zero-field $f=0$ given by (\ref{w+fsmall}).

\subsection{ Non-linear fluctuation theorem}

The non-linear fluctuation theorem obtained in \cite{c_nonlinearft}
says that, in any given sample of the trap model
the diffusion front $P_t^{(+f)}(n)$ in the presence of 
the external field $(+f)$ 
and the diffusion front $P_t^{(-f)}(n)$ in the presence 
of $(-f)$ are related by the extremely simple property
\begin{eqnarray}
\frac{ P_t^{(+f)}(n) }{ P_t^{(-f)}(n)} = e^{ \beta f n }
\label{ratio+-}
\end{eqnarray}
In the effective dynamics, this property is satisfied since 
the renormalization landscape at time $t$ is the same for $(+f)$ and $(-f)$,
so the two possible positions $M_+$ and $M_-$ are the same,
and the weights (\ref{formw+},\ref{formw-}) satisfy the properties
\begin{eqnarray}
 W_+(-F,  \lambda^+, \lambda^-) &&  = e^{-F \lambda_+}  W_+(F,  \lambda^+, \lambda^-)
\\
 W_-(-F,  \lambda^+, \lambda^-) &&  = e^{F \lambda_-}  W_-(F,  \lambda^+, \lambda^-)
\end{eqnarray}

\section{ Observables when the field is applied from the beginning $t \geq 0$}

\label{observablest0}

\subsection{ Diffusion front in a given sample }

In the effective dynamics, 
the diffusion front in a given sample takes the scaling form
\begin{eqnarray}
 P^{(0)}_{t}(n;f) = \frac{1}{\xi(t,f) } {\cal P}^{(0)} \left( X= \frac{n}{\xi(t,f) }  ; F= \beta f \xi(t,f) \right)
\end{eqnarray}
where the characteristic length scale $\xi(t,f)$
has been defined in (\ref{choicet},\ref{xit})
from the renormalization landscape, and where
the scaling function reads 
according to the effective dynamics (\ref{rulespplusmoins})
\begin{eqnarray}
 {\cal P}^{(0)} \left( X ,F \right) = 
\frac{ e^{-F \lambda^-} (1-e^{-F  \lambda^+} ) }
{ 1- e^{-F ( \lambda^++\lambda^-) }} \delta(X+\lambda^-)
+ \frac{1-e^{-F \lambda^-} }
{ 1- e^{-F ( \lambda^++\lambda^-) }} \delta(X- \lambda^+)
\label{pMM}
\end{eqnarray}
in terms of the two rescaled distances $\lambda_{\pm}$ between the origin and the nearest renormalized traps.
The joint distribution of the two
rescaled distances is completely factorized 
 \begin{eqnarray}
{\cal D}(  \lambda^+,\lambda^-) = \theta( \lambda^+) \theta(\lambda^-) 
e^{- \lambda^+ -\lambda^-} 
\label{measure0}
\end{eqnarray}

We now compute the various observables that can be obtained
from the sample-dependent diffusion front (\ref{pMM}) and 
the measure (\ref{measure0}) over the samples.

\subsection{ Disorder averaged diffusion front}

The averaged diffusion front takes the scaling form
\begin{eqnarray}
\overline{ P_t(n,f) }
 = \frac{1}{\xi(t,f)} 
g_{\mu} \left( X= \frac{n}{\xi(t,f)} , F= \beta f\xi(t,f)  \right)  
\end{eqnarray}
where the scaling function $g_{\mu}$ reads at lowest order in $\mu \to 0$ 
\begin{eqnarray}
g_0(X,F) 
&& = e^{- \vert X \vert }\left[ \theta(X>0)   + \theta(X<0) 
e^{- F \vert X \vert } \right]
 \int_0^{+\infty} d \lambda
  e^{ - \lambda} 
\frac{1-e^{-F \lambda} }
{ 1- e^{-F ( \vert X \vert+ \lambda) }} \\
&& = e^{- \vert X \vert }\left[ \theta(X>0)   + \theta(X<0) 
e^{- F \vert X \vert } \right]
\sum_{n=0}^{+\infty} \frac{F}{ (1+F n) (1+F+F n)}
e^{-n F \vert X \vert } 
\end{eqnarray}
which interpolates between the diffusion fronts
for the unbiased case \cite{c_agingtrap} as $F \to 0$
and for the directed case \cite{c_directed} as $F \to \infty$.
Here, as for other results below,
we have given two equivalent forms for the diffusion front : the first one with the integral is more appropriate to study the small $F$ behavior, whereas
the second one with the infinite series is more suited for large $F$.

The simple property 
\begin{eqnarray}
g_0(- X,F) = e^{- F  X  } g_0(X,F)
\end{eqnarray}
is actually expected to be valid for arbitrary $\mu$
as a consequence of a non-linear Fluctuation Theorem,
as discussed in \cite{c_nonlinearft}

\subsection{ Disorder averaged mean position }

The disorder average of the mean position 
takes the scaling form
\begin{eqnarray}
\overline{ <x(t,f)> } \equiv  \sum_{x=-\infty}^{+\infty}
 x  \overline{ P_t(x,f) }
= \xi(t,f) {\cal X}_{\mu} (F= \beta f\xi(t,f) ) 
\end{eqnarray}
where the scaling function $ {\cal X}_{\mu}$
 reads at lowest order in $\mu \to 0$ 
\begin{eqnarray}
{\cal X}_{0}(F)  \equiv \int_{-\infty}^{+\infty} dX X
g_0(X,F)  = \int_{0}^{+\infty} d\lambda \lambda
e^{-  \lambda  } \frac{ \frac{ F \lambda}{2} \coth \frac{ F \lambda}{2} -1}{ F} = 1- \frac{1}{F}-\frac{1}{F^3} \psi'' \left( 1+\frac{1}{F} \right)  
\label{calx0f}
\end{eqnarray}
The series expansion for small $F$ reads
\begin{eqnarray}
{\cal X}_{0}(F) && = \frac{F}{2} - \frac{F^3}{6}  + \frac{F^5}{6}+O(F^7)
\label{xofsmall}
\end{eqnarray}
whereas the asymptotic behavior for large $F$ reads
\begin{eqnarray}
{\cal X}_{0}(F) && = 1- \frac{1}{F}  +  \sum_{n=1}^{+\infty} \frac{2}{(1+F n)^3} =  1- \frac{1}{F} + \frac{2\zeta(3)}{F^3}- \frac{\pi^4}{15 F^4}
+O\left( \frac{1}{ F^5} \right) 
\label{xofbig}
\end{eqnarray}

Using $F= \beta f \xi(t,f)$ and (\ref{xifsmall},\ref{xifbig}), we finally
obtain the expansions
\begin{eqnarray}
\overline{ <x(t,f)> } && = \frac{\beta f}{2} 
\left(  \frac{t}{T_0} \right)^{\frac{ \mu}{1+\mu}}
 \left[ 
1 - \frac{1 }{3} (\beta f)^2  \left(  \frac{t}{T_0} \right)^{\frac{2\mu}{1+\mu}} + O( (\beta f t^{\frac{ \mu}{1+\mu}})^4)\right]
\label{moyxfsmall} 
\end{eqnarray}
at short times, i.e. for $t \ll t_{\mu}(f)$ (\ref{tmuf}), and
\begin{eqnarray}
\overline{ <x(t,f)> } = \left( \frac{\beta f}{2}\right)^{\mu} t^{\mu} 
 -   \frac{1}{\beta f} + O\left( \frac{1}{ (\beta f)^{2+\mu} t^{\mu})} \right) 
\label{moyxfbig} 
\end{eqnarray}
for $t \gg t_{\mu}(f)$.
The leading terms are thus in agreement with the scaling analysis presented in \cite{bertinjpreponse}.

\subsection{ Thermal width }

The disorder averaged thermal width (\ref{pMM}) presents the scaling form
\begin{eqnarray}
\overline{ <\Delta x^2(t,f)> }  \equiv 
\overline{ \sum_{x=-\infty}^{+\infty}
 x^2  P_t(x,f) - \left[ \sum_{x=-\infty}^{+\infty}
 x  P_t(x,f) \right]^2 } 
 = \xi^2(t,f) \Delta_{\mu}(F=\beta f \xi(t,f)) 
\label{defthermalwidtf} 
\end{eqnarray}
where the scaling function reads at lowest order in $\mu$

\begin{eqnarray}
\Delta_0(F)  = \int_0^{+\infty} d \lambda  e^{-\lambda} 
\frac{\lambda^2}{ 4 H } \left[ \coth \frac{ F \lambda}{2} 
- \frac{\frac{ F \lambda}{2}}{\sinh^2 \frac{ F \lambda}{2}  } \right]
 = \frac{1}{F} + \frac{2}{F^4} \psi'' \left( 1+\frac{1}{F} \right)
+ \frac{1}{F^5} \psi''' \left( 1+\frac{1}{F} \right)
\label{resdeltaof}
\end{eqnarray}
It is thus directly related to the scaling
function ${\cal X}_0(F)$ 
 governing the mean displacement (\ref{calx0f}) via
the simple relation
\begin{eqnarray}
\Delta_0(F) && = \left[ \frac{1}{F} + \frac{d}{d F}  \right]
{\cal X}_0(F)
\label{simpledeltafrommean}
\end{eqnarray}

Using the asymptotic behaviors for small and large $F$
\begin{eqnarray}
\Delta_0(F) && = 1 -\frac{2}{3} F^2 + F^4 +O(F^6) \\
\Delta_0(F) && = \frac{1}{F} - \frac{ 4 \zeta(3)}{F^4}+ \frac{\pi^4}{5 F^5}
+O\left( \frac{1}{ F^6} \right) 
\end{eqnarray}
with $F= \beta f \xi(t,f)$ (\ref{xifsmall}, \ref{xifbig}),
we obtain the following leading
terms for the unrescaled thermal width
\begin{eqnarray}
\overline{ <\Delta x^2(t,f)> } 
 =   \left(  \frac{t}{T_0} \right)^{\frac{2 \mu}{1+\mu}}
 \left[ 
1 - \frac{2}{3} (\beta f)^2 \left(  \frac{t}{T_0} \right)^{\frac{2\mu}{1+\mu}} + O( (\beta f)^4 t^{\frac{4\mu}{1+\mu}})\right] 
\end{eqnarray}
for $t\ll t_{\mu}(f)$ (\ref{tmuf}), and
\begin{eqnarray}
\overline{ <\Delta x^2(t,f)> } 
=  \frac{\left( \frac{\beta f}{2} \right)^{\mu} t^{\mu} }{ \beta f} 
\left[ 1  - \frac{ 4 \zeta(3)}{ [ (\beta f)^{1+\mu} t^{\mu} ]^3 } 
+ O\left( \frac{ 1}{ [ (\beta f)^{1+\mu} t^{\mu} ]^4 } \right) \right]
\label{deltax2tlarge} 
   \end{eqnarray}
for $t \gg t_{\mu}(f)$.

\subsection{ Mean square displacement }

In addition to the thermal width (\ref{defthermalwidtf}), 
it is also interesting to consider separetely
the two terms, namely 
the mean square displacement 
\begin{eqnarray}
{\cal D}_0(F) \equiv \overline{ < X^2> }
 = 2-\frac{2}{F} +\frac{2}{F^2}+ \frac{2}{F^4} \psi'' \left( 1+\frac{1}{F} \right) =  1+\frac{F^2}{3} -\frac{F^4}{3} +O(F^6)
\label{cald0f}
\end{eqnarray}
and the sqare of the mean displacement
\begin{eqnarray}
\overline{< X>^2}
 = 2-\frac{3}{F} +\frac{2}{F^2}- \frac{1}{F^5} \psi''' \left( 1+\frac{1}{F} \right) = F^2  -\frac{4 F^4}{3} +O(F^6)
\end{eqnarray}
in the rescaled variable $X=x/\xi(t,f)$ as usual.

\subsection{ Localization parameters }

The localization parameters, defined as 
the disorder averages of the probabilities
to find $k$ independent particle in the same trap,
follow the scaling form
\begin{eqnarray}
\overline{ Y_k^{(\mu)}(t,f) } \equiv
 \sum_{x=-\infty}^{+\infty} \overline{ [ P_t^{(f)}(x) ]^k} = {\cal Y}_k^{(\mu)}(F=\beta f \xi(t,f))
\label{defyk}
   \end{eqnarray}
where the scaling function at lowest order in $\mu$ 
reads (\ref{pMM})
\begin{eqnarray}
 {\cal Y}_k^{(0)}(F)
 && \equiv \int_0^{+\infty} d \lambda^+ \int_0^{+\infty} d\lambda^-
e^{- \lambda^+ -\lambda^-}
\left(  \left[ \frac{ e^{-F \lambda^-} (1-e^{-F  \lambda^+} ) }
{ 1- e^{-F ( \lambda^++\lambda^-) }} \right]^k
+ \left[ \frac{1-e^{-F \lambda^-} }
{ 1- e^{-F ( \lambda^++\lambda^-) }} \right]^k  \right) 
\nonumber  \\
&& = \sum_{m=0}^{+\infty}
\frac{ k \Gamma \left( \frac{1}{F} + m \right) \Gamma (k + m)  }
{ F \Gamma (1 + m) \Gamma \left( 1 + \frac{1}{F} + k + m] \right) }
\left[ \frac{1}{1 + F m} +\frac{1}{1 + F (k + m)} \right]
\label{resyk}
\end{eqnarray}

The series expansion for small $F$ gives the first
correction with respect to the unbiased case \cite{c_agingtrap}
\begin{eqnarray}
 {\cal Y}_k^{(\mu)}(F)
 && = \frac{2}{k+1} + F^2 \frac{2 k (k-1)}{ (k+1)(k+2)(k+3)}
+F^4 \frac{2 k(k-1)(k^2-15k-4)  }
{(k+1 ) (k+2 ) (k+3 ) (k+4 ) (k+5 ) } +O(F^6)
\end{eqnarray}
whereas for large $F$, the first corrections with the directed case
\cite{c_directed} read 
\begin{eqnarray}
 {\cal Y}_k^{(0)}(F)
 && = 1 - (\gamma_{Euler} + \psi(k)) \frac{1}{F}
+ \left( \frac{ (\gamma_{Euler} + \psi(k))^2 }{2} + \frac{\pi^2}{4} - \frac{3}{2} \psi'(k) \right) \frac{1}{F^2} +O \left( \frac{1}{F^3} \right)
\end{eqnarray}

For $k=2$ and $k=3$, we may moreover sum the series (\ref{resyk}) to obtain
\begin{eqnarray}
 {\cal Y}_2^{(0)}(F)
&& = 1 - \frac{1}{F}
+ \frac{2}{F^2} - \frac{2}{F^3} \psi' \left( 1+\frac{1}{F} \right)  \\
 {\cal Y}_3^{(0)}(F)
&& = 1 - \frac{3}{2 F}
 + \frac{3}{F^2} - \frac{3}{F^3} \psi' \left( 1+\frac{1}{F} \right)
\end{eqnarray}
which are thus related by the very simple relation
\begin{eqnarray}
\frac{1- {\cal Y}_2^{(0)}(F)}{2} = \frac{1- {\cal Y}_3^{(0)}(F)}{3}
\label{simplerelationk23}
\end{eqnarray}
This relation actually reflects the the two-delta structure 
of the diffusion front in any given sample :
indeed, if we note $p_1$ and $p_2$ the weights
of the two delta peaks with $p_1+p_2=1$, it is 
immediate to obtain the relation (\ref{simplerelationk23}) since 
\begin{eqnarray}
1- Y_2^{(0)} && = (p_1+p_2)^2-p_1^2-p_2^2= 2 p_1 p_2 \nonumber \\
1- Y_3^{(0)} && = (p_1+p_2)^3-p_1^3-p_2^2= 3 p_1 p_2 (p_1+p_2) = 3 p_1 p_2
\end{eqnarray}

\subsection{ Entropy }

The entropy (\ref{defentropy}) is closely related to 
the localization parameters (\ref{defyk}) since it reads
\begin{eqnarray}
\overline{ S(t,f) } \equiv - \sum_{x=-\infty}^{+\infty} 
\overline{ P_{t}(x)  \ln P_{t}(x) }
= - \left[ \frac{ \partial 
\overline{ Y_k^{(\mu)}(t,f)  }  }
{\partial k} \right]_{\vert_{k=1}}   = {\cal S}^{(\mu)}(F=\beta f \xi(t,f))
 \end{eqnarray}
where the scaling function reads at lowest order in $\mu$ 
\begin{eqnarray}
 {\cal S}^{(0)}(F) && =  \frac{\psi''\left(1+\frac{1}{F}\right)}{F^2}
+ \sum_{m=0}^{+\infty} \left( \frac{1}{(1+F+F m)^2}- \frac{1}{(1+F m)^2} \right)
\left[ \psi\left(1+m\right)-\psi\left(1+\frac{1}{F}+m\right)\right]
 \end{eqnarray}

The asymptotic behaviors are given by the series
\begin{eqnarray}
 {\cal S}^{(0)}(F)  = \frac{1}{2} - \frac{F^2}{12} + \frac{F^4}{20}
+O(F^6)
 \end{eqnarray}
for small $F$ and by 
\begin{eqnarray}
{\cal S}^{(0)}(F)  && = \frac{\pi^2}{6 F}
-  \frac{ 3 \zeta(3)}{F^2}+O\left(\frac{1}{F^3} \right)
\end{eqnarray}
 for large $F$.
So for the unbiased case $f=0$, the entropy
remains forever frozen at the value ${\cal S}^{(0)}(F=0)  = 1/2$.
For the biased case, the entropy decays towards zero 
as the directed character gets stronger as time grows.

\subsection{ Two-particle correlation function }

The two-particle correlation function reads
\begin{eqnarray}
 C(l,t) \equiv \sum_{x=-\infty}^{+\infty} \sum_{x'=-\infty}^{+\infty}
\overline{P_t(x) P_t(x') }\delta_{l,\vert x-x' \vert} 
\opsimeq_{t \to \infty} Y_2^{(0)} \delta_{l,0}
+ \frac{1}{\xi(t,f) } {\cal C}_{\mu} \left( \lambda= \frac{l}{\xi(t,f)} \right)
\end{eqnarray} 
where the weight of the delta peak corresponds as it should to the localization parameter $ Y_2^{(0)} $
discussed above, whereas the scaling function of the long-ranged part reads at lowest order
\begin{eqnarray}
  {\cal C}_0 \left( \lambda ,F \right) 
 = e^{-\lambda} \frac{ 1 - 2 F \lambda e^{-F \lambda}- e^{-2 F \lambda} }{ F (1-e^{-F \lambda})^2}    
\end{eqnarray} 

In particular, the first correction 
with respect to the unbiased case reads \cite{c_agingtrap}
\begin{eqnarray}
  {\cal C}_{\mu}^{(0)} \left( \lambda ,F  \right) 
 = e^{-\lambda}   \frac{\lambda}{3} \left[ 1 - \frac{ F^2 \lambda^2}{30} 
+ O(F^4 \lambda^4) \right] 
\end{eqnarray}

\section{ Effective dynamics when the field $f$ is applied for $t \geq t_w$}

\label{dynamicsfttw}

We now study the dynamics in the following aging ``experiment" 
already considered in \cite{bertinjpreponse}:
the system first evolves with no external field $f=0$ during the interval $t \in [0,t_w]$, and then an external field $f>0$ is applied for $t \in [t_w,+\infty[$.

\subsection{ Time sector governed by the effective dynamics  }

The scale of the renormalized landscape
corresponding to time $t_w$ and to no external field
$f=0$ reads (\ref{roftfsmall}) 
\begin{eqnarray}
R(t_w,f=0) = \left(  \frac{t_w}{T_0} \right)^{\frac{1}{1+\mu}} 
\label{defrw}
\end{eqnarray}
The corresponding length scale between renormalized traps
is given by (\ref{xifsmall}) 
\begin{eqnarray}
\xi(t_w,f=0) = \left(  \frac{t_w}{T_0} \right)^{\frac{\mu}{1+\mu}} 
\end{eqnarray}

The state reached at $t_w$, which is made out 
of two delta peaks (\ref{pMM}),
 has to be considered as an
initial condition for the dynamics in the presence of $f>0$
in the new time $(t-t_w)$.
Since at $t_w$, the particle is typically in a trap of trapping time $\tau>R(t_w,f=0)$, the effective dynamics corresponds to
no move as long as $R(t-t_w,f)<R(t_w,0)$, whereas
the effective dynamics governed by the decimation procedure   
becomes ``active" for $R(t-t_w,f)>R(t_w,f=0)$.
As a consequence, it is useful to introduce the parameter
\begin{eqnarray} 
\alpha(t,t_w,f) \equiv \frac{ \xi(t-t_w,f) }{ \xi(t_w,f=0)}
= \frac{ R^{\mu}(t-t_w,f) }{  R^{\mu}(t_w,f=0)}
\label{defalpha}
\end{eqnarray}
that measures the ratio of the length scales of
renormalized landscape at the two scales $ R(t-t_w,f)$ and $R(t_w,f=0)$.
The response in the time sector $\alpha<1$,
which is governed by rare events
will be discussed separately in the section \ref{rareevents}.
In this Section, we will concentrate on the time sector
$\alpha(t,t_w,f)>1$, where
the effective dynamics governed by the decimation procedure   
is the leading effect.

\subsection{ Role of the characteristic time $t_{\mu}(f)$
associted to the external field  }

We have already seen that a bias $f$ introduces the characteristic time scale 
$t_{\mu}(f)= \left( \frac{1}{\beta f }\right)^{\frac{1+\mu}{\mu}}$
(\ref{tmuf}). It is thus natural that the domain in $(t-t_w)$
corresponding to the time sector $\alpha>1$
actually depends on the relative value of $t_w$
with respect to $t_{\mu}(f)$.

For $t_w \ll t_{\mu}(f)$, 
the time sector $\alpha>1$ corresponds to the domain
$(t-t_w)>t_w$, and the ratio $\alpha$ interpolates between
\begin{eqnarray} 
\alpha(t,t_w,f) && =  
\left( \frac{ t-t_w }{ t_w} \right)^{\frac{\mu}{1+\mu}}
\ \ \rm{for} \ \  t_w < t-t_w \ll   t_{\mu}(f)
\label{ashort} 
\end{eqnarray}
and
\begin{eqnarray} 
\alpha(t,t_w,f) && =  
\left( \frac{  \frac{\beta f}{2}   (t-t_w) }
{ t_w^{\frac{1}{1+\mu}}} \right)^{\mu}
\ \ \rm{for} \ \   t-t_w \gg   t_{\mu}(f)
\label{along}
\end{eqnarray}

On the other hand, for $t_w \gg t_{\mu}(f)$, the 
the time sector $\alpha>1$ corresponds to the domain
$(t-t_w)>\frac{2}{\beta f} t_w^{\frac{1}{1+\mu}}$, and we have (\ref{along})
in the whole time sector $\alpha>1$.

To simplify the reading in the following, we will use the simplified
notations
\begin{eqnarray} 
R && \equiv R(t-t_w,f) \nonumber \\
R_w && \equiv R(t_w,f=0) \nonumber \\
\alpha && \equiv \alpha(t,t_w,f) = \frac{R^{\mu}} {R_w^{\mu}}
\end{eqnarray}

\subsection{ Statistical properties of the renormalized landscape
at two successive scales}

To study the two-time effective dynamics,
we will need the probability measure for the renormalized landscape
at the two successive scales $R_w$ and $R$. Since the two sides
of the origin are independent, we first consider the half line $x>0$
alone.
The joint probability that the first renormalized trap at scale $R_w$
is at distance $l_w^+$, has a trapping time $\tau_w^+$ and that the first
renormalized trap at scale $R$ is at distance $l^+$ reads 
\begin{eqnarray} 
 A_{R_w,R}(\tau_w^+,l_w^+;l^+) =
 \theta(\tau_w^+>R_w) \frac{\mu}{(\tau_w^+)^{1+\mu} } 
e^{- \frac{l_w^+}{R_w^{\mu}}}      \left[ \theta(\tau_w^+>R) 
\delta(l^+-l_w^+)
+ \theta(\tau_w^+<R) \theta(l^+>l_w^+) 
\frac{1}{R^{\mu}} e^{- \frac{(l-l_w^+)}{R^{\mu}}  }
  \right]
\end{eqnarray}
After integration over the trapping time
$\tau_w^+$, we obtain the scaling form
\begin{eqnarray} 
 A_{R_w,R}(l_w^+;l^+) \equiv \int d\tau_w^+ A_{R_w,R}(\tau_w^+,l_w^+;l^+)
= \frac{1}{R_w^{\mu} R^{\mu}}  {\cal A}
 \left( \lambda_w^+ \equiv \frac{ l_w^+}{ R_w^{\mu} }
;  \lambda^+ \equiv \frac{ l^+-l_w^+}{ R^{\mu}  } ;
\alpha \equiv \frac{ R^{\mu}}{ R_w^{\mu} } \right)
\label{ascalingform}
\end{eqnarray}
with the scaling function
\begin{eqnarray} 
{\cal A} \left( \lambda_w^+  
;  \lambda^+ ;\alpha\right)
=    
e^{- \lambda_w^+ }       
\left[  \frac{1}{\alpha}  \delta(\lambda^+)
+ \left( 1-  \frac{1}{\alpha}   \right) \theta(\lambda^+>0) 
 e^{-  \lambda^+ }
   \right]
\label{arescaled}
\end{eqnarray}

Since we have the same properties for the half-line $x<0$,
the measure for the full line reads

\begin{eqnarray} 
&& {\cal A} \left( \lambda_w^+; \lambda_w^-  
;  \lambda^+; \lambda^- ;\alpha\right) \\
&& = \theta(\lambda^+_w>0) \theta(\lambda^-_w>0) e^{- \lambda_w^+- \lambda_w^- }       \left[  \frac{1}{\alpha}  \delta(\lambda^+)
+ \left( 1-  \frac{1}{\alpha}   \right) \theta(\lambda^+>0) 
 e^{-  \lambda^+ }   \right]
\left[  \frac{1}{\alpha}  \delta(\lambda^-)
+ \left( 1-  \frac{1}{\alpha}   \right) \theta(\lambda^->0) 
 e^{-  \lambda^- }   \right] \nonumber
\label{atot}
\end{eqnarray}

\subsection{ Rules for the two-time effective dynamics }

Given the configuration $(\lambda^-,\lambda_w^-,\lambda_w^+,\lambda^+)$ charaterising
the renormalized landscape at the two successive scales, 
the two-time diffusion front takes the scaling form
\begin{eqnarray} 
 P(x,t ; x_w,t_w \vert 0,0)  
  = \frac{1}{ R^{\mu}_w R^{\mu}} 
 {\cal P}^{(\mu)}_{\{ \lambda^-,\lambda_w^-,\lambda_w^+,\lambda^+ \}}
 \left( X_w = \frac{ x_w }{ R^{\mu}_w }  
, Y = \frac{ x-x_w }{ R^{\mu} } , F =
\beta f R^{\mu}, \alpha = \frac{R^{\mu}}{R^{\mu}_w} \right)
\end{eqnarray}
with
\begin{eqnarray} 
&& {\cal P}^{(0)}_{\{ \lambda^-,\lambda_w^-,\lambda_w^+,\lambda^+ \}}
 \left( X_w , Y  , F, \alpha \right)
 = \\
&&  \frac{\lambda_w^- }{\lambda_w^+ + \lambda_w^-}  \delta(X_w-\lambda_w^+) 
\nonumber \\
&& \times \left[  W_+(\lambda^+, \lambda^- +\frac{\lambda_w^+ + \lambda_w^-}{\alpha} ,F) \delta(Y-\lambda^+)
+ W_-(\lambda^+, \lambda^- +\frac{\lambda_w^+ + \lambda_w^-}{\alpha} ,F)  \delta(Y+\lambda^- +\frac{\lambda_w^+ + \lambda_w^-}{\alpha} )\right] 
\nonumber  \\
&& + \frac{\lambda_w^+ }{\lambda_w^+ + \lambda_w^-} \delta(X_w+\lambda_w^-) 
\nonumber \\
&& \times \left[  W_+(\lambda^+ +\frac{\lambda_w^+ + \lambda_w^-}{\alpha},\lambda^-,F) \delta(Y- \lambda^+ - \frac{\lambda_w^+ + \lambda_w^-}{\alpha})
+  W_-(\lambda^+ +\frac{\lambda_w^+ + \lambda_w^-}{\alpha},\lambda^-,F)  \delta(Y+ \lambda^-)\right] \nonumber
\label{ptwotimeinasample}
\end{eqnarray}

We are now in position to compute various observables in the aging regime
from the knowledge of the sample-dependent two-time diffusion front
(\ref{ptwotimeinasample}) and from the measure (\ref{atot}) over the samples.

\section{ Observables when the field $f$ is applied for $t \geq t_w$}

\label{observablesttw}

\subsection{ Disorder averaged two-time diffusion front  }

The disorder averaged diffusion front with
the measure (\ref{atot}) reads
\begin{eqnarray} 
&& \overline { {\cal P}^{(0)} \left( X_w  , Y , F, \alpha \right) }
\equiv  \int_0^{+\infty} d\lambda_w^+ \int_0^{+\infty} d\lambda_w^-
\int_0^{+\infty} d\lambda^+ \int_0^{+\infty} d\lambda^- 
{\cal A} \left( \lambda_w^+; \lambda_w^-  
;  \lambda^+; \lambda^- ;\alpha\right) {\cal P}^{(0)}_{\{ \lambda^-,\lambda_w^-,\lambda_w^+,\lambda^+ \}}
 \left( X_w , Y  , F, \alpha \right) \nonumber \\  && = 
\left[\theta(X_w \geq 0) \theta(Y \geq 0 ) +
e^{-F \vert Y \vert } \theta(X_w \leq 0) \theta(Y\leq 0 ) \right]
 g^{(0)}_{++}\left( \vert X_w \vert  , \vert Y \vert, F, \alpha \right) \\
&& + \left[\theta(X_w \leq 0) \theta(Y \geq 0 ) +
e^{-F \vert Y \vert } \theta(X_w \geq 0) \theta(Y \leq 0 ) \right]
 g^{(0)}_{-+}\left( \vert X_w \vert  , \vert Y \vert , F, \alpha \right)
\label{jointxwy}
\end{eqnarray}
where
\begin{eqnarray} 
g^{(0)}_{++}\left( X_w  , Y , F, \alpha \right)
&& = e^{- X_w } \left[  \frac{1}{\alpha}  \delta(Y)
+ \left( 1-  \frac{1}{\alpha}   \right)   e^{-  Y }   \right]
  \int_0^{+\infty} d\lambda_w^- e^{- \lambda_w^- } 
\frac{\lambda_w^- }{X_w + \lambda_w^-}  \\  
&&    \left[  \left( \frac{1}{\alpha} \right) 
\frac{1-e^{-F \left( \frac{X_w + \lambda_w^-}{\alpha} \right) } }
{ 1-e^{-F \left( Y+ \frac{X_w + \lambda_w^-}{\alpha} \right) }}
+ \left( 1-  \frac{1}{\alpha}   \right)  
\int_0^{+\infty} d\lambda^- e^{-  \lambda^- } 
\frac{1-e^{-F \left( \lambda^- +\frac{X_w + \lambda_w^-}{\alpha} \right) } }
{ 1-e^{-F \left( Y+ \lambda^- +\frac{X_w + \lambda_w^-}{\alpha} \right) }}  \right] 
\label{g++}
\end{eqnarray}
and
\begin{eqnarray} 
&& g^{(0)}_{-+}\left(  X_w  , Y , F, \alpha \right)
 =  e^{-  X_w  } \left( 1-  \frac{1}{\alpha}   \right)
 \int_0^{+\infty} d\lambda^-   e^{-  \lambda^- }   
 \frac{1-e^{-F \lambda^- } }{ 1-e^{-F (Y+\lambda^-) } } \\
&& \int_0^{+\infty} d\lambda_w^+  e^{- \lambda_w^+ } 
 \frac{\lambda_w^+ }{\lambda_w^+ +  X_w } 
      \left[  \frac{1}{\alpha}  \delta(Y-  \frac{\lambda_w^+ +  X_w }{\alpha}) 
+ \left( 1-  \frac{1}{\alpha}   \right) 
\theta( Y-  \frac{\lambda_w^+ +  X_w }{\alpha})  
 e^{-  ( Y-  \frac{\lambda_w^+ +  X_w }{\alpha})  }   \right]  
\label{g-+}  
\end{eqnarray}

The form (\ref{jointxwy}) presents the simple property
\begin{eqnarray} 
\overline { {\cal P}^{(0)} \left( -X_w  , -Y , F, \alpha \right) }
= e^{- F Y} \overline { {\cal P}^{(0)} \left( X_w  , Y , F, \alpha \right) }
\end{eqnarray}
which is expected to be true for arbitrary $\mu$, as a consequence
of the non-linear Fluctuation Theorem discussed in \cite{c_nonlinearft}.

\subsection{ Law of the relative displacement $Y$ between $t_w$ and $t$}

We now consider the partial law of the rescaled relative displacement $Y$
between $t_w$ and $t$, i.e. we integrate 
the disorder averaged diffusion front
(\ref{jointxwy}) over the position $X_w$ at time $t_w$ : we obtain the form
\begin{eqnarray} 
 \overline { {\cal P}^{(0)} \left( Y , F, \alpha \right) }
  \equiv \int_{-\infty}^{+\infty} dX_w 
 \overline { {\cal P}^{(0)} \left( X_w  , Y , F, \alpha \right) }  = 
\left[ \theta(Y \geq 0 ) +
e^{-F \vert Y \vert }  \theta(Y \leq 0 ) \right] G \left(  Y , F, \alpha \right)
\end{eqnarray}
where
\begin{eqnarray} 
&&  G \left(  Y , F, \alpha \right) =
\int_0^{+\infty} dX_w 
\left[ g^{(0)}_{++}\left( \vert X_w \vert  , \vert Y \vert, F, \alpha \right)
+  g^{(0)}_{-+}\left( \vert X_w \vert  , \vert Y \vert , F, \alpha \right)
\right]
\end{eqnarray}

  Using the intermediate results
  \begin{eqnarray} 
  && \int_0^{+\infty} dX_w  g^{(0)}_{++}\left( X_w  , Y , F, \alpha \right)
  =    \frac{1}{2 \alpha}  \delta(Y)
  +   \frac{1}{2}   e^{-  Y }  \int_0^{+\infty} du \frac{1-e^{-F u } }
  { 1-e^{-F \left( Y+ u\right) }}    
    \left[ e^{-u} -  e^{- \alpha  u } \right]  
  \end{eqnarray}
and
\begin{eqnarray} 
&& \int_0^{+\infty} dX_w 
 g^{(0)}_{-+}\left( \vert X_w \vert  , \vert Y \vert, F, \alpha \right) 
=   \frac{ 1 }{2} ( e^{-Y} - e^{-  \alpha Y  } ) 
 \int_0^{+\infty} du   e^{-  u }   
 \frac{1-e^{-F u } }{ 1-e^{-F (Y+u) } }
  \end{eqnarray}
we finally obtain
\begin{eqnarray}  
 \overline { {\cal P}^{(0)} \left(  Y , F, \alpha \right) }  = 
 \frac{1}{ \alpha}  \delta(Y)
+ \left[ \theta(Y \geq 0 ) +
e^{-F \vert Y \vert }  \theta(Y \leq 0 ) \right] 
G_{ns} \left( \vert Y \vert, F, \alpha \right)
\label{reslawy}
\end{eqnarray}
where the non-singular part is given in terms of the function
\begin{eqnarray}  
 G_{ns} \left(  Y , F, \alpha \right)
&& =  \frac{e^{-Y}}{2}    \int_0^{+\infty} du \frac{1-e^{-F u } }
{ 1-e^{-F \left( Y+ u\right) }}    
\left[ ( 2 -  e^{-  (\alpha-1) Y  } )  e^{-  u }
   -  e^{- \alpha  u }    \right] \\
&& =  e^{-Y} \sum_{n=0}^{+\infty} e^{ -Fn Y} 
\left[ \frac{H}{(1+F n) (1+F+F n) } (2-e^{- (\alpha-1) Y} ) - \frac{H}{(a+F n) (a+F+F n) }
 \right]
\end{eqnarray}

The presence of a singular part in $\delta(Y)$ for the 
rescaled relative displacement 
was already found in the case of the Sinai model
by the RSRG approach \cite{us_sinai}
and by mathematicians \cite{deltaymath}.
Here the weight of $\delta(Y)$ is simply given by 
$(1/\alpha)$ (\ref{defalpha})
which represents the probability for a renormalized trap
at scale $\xi(f=0,t_w)$ to be still present
in the renormalized landscape at the new scale $\xi(f,t-t_w)$.

\subsection{ Disorder averaged mean position }

Since $<x_w>=0$ in any sample, in the effective dynamics
in the limit $\mu \to 0$ but also more generally
for any $\mu$
as a consequence of a special dynamical symmetry \cite{c_nonlinearft},
the mean position $<x>$ at time $t$ in a sample
also corresponds to the mean displacement $<(x-x_w)>$
between the times $t_w$ and $t$.
The disorder average of the mean position
takes the scaling form the scaling form 
\begin{eqnarray}  
\overline{ <x>(t,t_w,f) }
= R^{\mu} {\cal X}_{\mu}(F =
\beta f R^{\mu}, \alpha = \frac{R^{\mu}}{R^{\mu}_w}  )
\label{defxttwf}
\end{eqnarray}
where the scaling function 
is simply given in terms of the law of the relative displacement $Y$
(\ref{reslawy})  
\begin{eqnarray}  
 {\cal X}_{0}(F , \alpha  )
&& = \frac{1}{2}  \int_0^{+\infty} dv  \frac{ 1 } 
{ 1-e^{-F v }}
 \int_{0}^{v} dY Y e^{-Y}
( 1- e^{-F  Y  }  )   
  (1-e^{F Y-F v })   
\left[ ( 2 -  e^{-  (\alpha-1) Y  } )  e^{Y- v  }
   -  e^{ \alpha Y - \alpha  v }    \right] \\
&& = 1 - \frac{1}{F}  
-\frac{1}{F^3} \psi'' \left( 1+\frac{1}{F} \right) 
- \frac{1}{2 \alpha} - \frac{1}{2 \alpha^2}
+ \frac{1}{ 2 (\alpha-1+F)} - \frac{1}{ 2 \alpha^2 (\alpha-1-F)} \\
&& 
+ \frac{1}{(\alpha-1) ((\alpha-1)^2- F^2)} 
\left[ \psi' \left( 1+\frac{1}{F} \right)
- \psi' \left( 1+\frac{\alpha}{F} \right)\right]
\label{resxfa}
\end{eqnarray}
We now consider various limit expressions.

On one hand, for fixed $\alpha$, the asymptotic expressions
for small $F$ and large $F$  
which generalize (\ref{xofsmall},\ref{xofbig}) read
\begin{eqnarray}  
 {\cal X}_{0}(F , \alpha  )
= \left[ 1 - \frac{1}{\alpha^2} \right] \frac{F}{2}
 - \frac{1}{6}
\left[ 1 - \frac{1}{\alpha^3} \right] F^3 
+ \frac{1}{6}
\left[ 1 - \frac{1}{5} 
\left( \frac{1}{\alpha^3} + \frac{3}{\alpha^4}+\frac{1}{\alpha^5} \right) \right] F^5 +O(F^7) 
\label{xfsmall}
\end{eqnarray}
and
\begin{eqnarray}  
 {\cal X}_{0}(F , \alpha  )
= \left( 1 - \frac{1}{2 \alpha} - \frac{1}{2 \alpha^2}  \right)
- \frac{1}{2} \left( 1  - \frac{1}{ \alpha^2}  \right) \frac{1}{F} 
- \frac{1}{2} \left( \alpha-1- \frac{1}{ \alpha} + \frac{1}{ \alpha^2}  \right) \frac{1}{F^2}
 +O(\frac{1}{F^3}) 
\label{xfbig}
\end{eqnarray}

On the other hand, for fixed $F$,
we have at the beginning of the aging regime where $(\alpha-1)$
is small
\begin{eqnarray}  
 {\cal X}_{0}(F , \alpha  )
&& = (\alpha-1) 
 \left[ \frac{3}{2} - \frac{1}{F} + \frac{1}{2 F^4}
\psi''' \left( 1+\frac{1}{F} \right)  \right]
 +O( (\alpha-1)^2) 
\label{xashort}
\end{eqnarray}
whereas for large $\alpha$,  
the first corrections with respect to the previous result (\ref{calx0f})
corresponding to $\alpha \to \infty$ read
\begin{eqnarray}  
 {\cal X}_{0}(F , \alpha  )
= 1- \frac{1}{F}-\frac{1}{F^3} \psi'' \left( 1+\frac{1}{F} \right)  
- \frac{F }{2 \alpha^2}
+ \left[ -F + \frac{F^2}{2} + \psi' \left( 1+\frac{1}{F} \right) \right] \frac{1}{\alpha^3} +O(\frac{1}{\alpha^4})
\label{xalong}
\end{eqnarray}

\subsection{ Mean-sqare displacement during $[t_w,t]$ }

Similarly, the mean-sqare displacement during $[t_w,t]$
takes the scaling form
\begin{eqnarray} 
D(t,t_w) = \overline{ <(x(t)- x(t_w))^2>  }
 =  R^{2\mu} {\cal D}_{\mu}(F =
\beta f R^{\mu}, \alpha = \frac{R^{\mu}}{R_w^{\mu}}  )
\end{eqnarray}
where the scaling function 
is again obtained from the law of the relative displacement $Y$
(\ref{reslawy})
\begin{eqnarray}  
&& {\cal D}_{0}(F , \alpha  )
 = 
 \int_{0}^{+\infty} dY Y^2 
\left[ 1+ e^{-F  Y  }   \right] \frac{e^{-Y}}{2}    \int_0^{+\infty} du \frac{1-e^{-F u } }
{ 1-e^{-F \left( Y+ u\right) }}    
\left[ ( 2 -  e^{-  (\alpha-1) Y  } )  e^{-  u }
   -  e^{- \alpha  u }    \right] \\
&& = 2 -\frac{2}{F} + \frac{2}{F^2}
-\frac{1}{\alpha}-\frac{1}{\alpha^3}
+ \frac{1}{\alpha-1+F}-\frac{1}{(\alpha-1+F)^2}
-\frac{1}{\alpha^2 (\alpha-1-F)^2}
-\frac{1}{\alpha^3 (\alpha-1-F)} \\
&& + \frac{ 4 (\alpha-1) }{ F ((\alpha-1)^2 -F^2) }
\left[ \psi'\left( 1+\frac{1}{F} \right) -  \psi'\left( 1+\frac{\alpha}{F} \right) \right]
\\
&& + \frac{2 (\alpha-1)^2 -F^2 }{ F^4 ((\alpha-1)^2 -F^2) }
 \psi''\left( 1+\frac{1}{F} \right)  + \frac{1 }{ F^2 ((\alpha-1)^2 -F^2) }
 \psi''\left( 1+\frac{\alpha}{F} \right) \nonumber 
\end{eqnarray}

For fixed $\alpha$, the asymptotic expressions
for small $F$ and large $F$ respectively read
\begin{eqnarray}  
 {\cal D}_{0}(F , \alpha  )
=  1-\frac{1}{\alpha^2} + \left[ \frac{1}{3}- \frac{1}{ 2 \alpha^2}
+ \frac{2}{3 \alpha^3}  - \frac{1}{ 2 \alpha^4} \right] F^2 +
\left[ - \frac{1}{3}
+\frac{1}{ 6 \alpha^2}
+ \frac{1}{5 \alpha^3}  - \frac{2}{ 5 \alpha^4} 
+\frac{1}{ 5 \alpha^5}
+ \frac{1}{6 \alpha^6}
 \right]F^4+O(F^6)
\end{eqnarray}
and
\begin{eqnarray}  
{\cal D}_{0}(F , \alpha  )
=  2 - \frac{1}{\alpha}  - \frac{1}{\alpha^3}
- \left(1 - \frac{1}{\alpha^3} \right) \frac{1}{F}
+ \left(2 - \alpha - \frac{1}{\alpha^3} \right) \frac{1}{F^2}
+O \left( \frac{1}{F^3} \right)
\end{eqnarray}

For fixed $F$, we have at the beginning of the aging regime where $(\alpha-1)$
\begin{eqnarray}  
&& {\cal D}_{0}(F , \alpha  )
=  \left[ 4 -\frac{3}{F} + \frac{2}{F^2} - \frac{1}{F^5}
\psi'''\left( 1+\frac{1}{F} \right) \right] (\alpha-1)
+O \left( (\alpha-1)^2 \right)
\end{eqnarray}
whereas for large $\alpha$, we obtain 
the first corrections to the previous result (\ref{cald0f})
\begin{eqnarray}  
 {\cal D}_{0}(F , \alpha  )
&& =  \left[ 2 -\frac{2}{F} + \frac{2}{F^2} + \frac{2}{F^4}
\psi''\left( 1+\frac{1}{F} \right) \right] 
+ \left[ - F + \frac{1}{F^2}
\psi''\left( 1+\frac{1}{F} \right) \right] \frac{1}{\alpha^2}
\\
&& + \left[ -2+ F^2 + \frac{4}{F}
\psi'\left( 1+\frac{1}{F} \right) + \frac{2}{F^2}
\psi''\left( 1+\frac{1}{F} \right) \right] \frac{1}{\alpha^3}
+O \left( \frac{1}{\alpha^4} \right)
\end{eqnarray}

\subsection{ Thermal width }

The rescaled thermal width 
\begin{eqnarray} 
\Delta_{\mu}(F,\alpha) = \overline{<X^2>-<X>^2}
\label{thermalwidthalpha}
\end{eqnarray}
may be similarly computed,
but since the full expression is rather lengthy, we will
only give the asymptotic forms.

For fixed $\alpha$, the asymptotic expressions
for small $F$ and large $F$ respectively read 
\begin{eqnarray}
\Delta_{0}(F,\alpha) && =  1 + 
\left[ -\frac{2}{3}
+ \frac{1}{ \alpha^3}  - \frac{1}{ 3 \alpha^4} \right] F^2 +
\left[ 1
- \frac{5}{3 \alpha^4}   
+\frac{4}{ 9 \alpha^5}
+ \frac{2}{9 \alpha^6}
 \right]F^4+O(F^6)
\end{eqnarray}
and 
\begin{eqnarray}  
\Delta_{0}(F,\alpha) && 
=  \frac{1}{3 \alpha} + \frac{1}{3 \alpha^2}+\frac{1}{3 \alpha^3}
+ \left( \frac{2}{3}    - \frac{2}{3\alpha^3} \right) \frac{1}{F}
+ \left(\frac{\alpha}{6} - \frac{1}{6}
- \frac{5}{6 \alpha^2} + \frac{5}{6\alpha^3}  \right) \frac{1}{F^2}
+O \left( \frac{1}{F^3} \right)
\end{eqnarray}

For fixed $F$,
at the beginning of the aging regime where $(\alpha-1)$ is small, we have
\begin{eqnarray}
\Delta_{0}(F,\alpha)
&& = 1+ \left[ -2 + \frac{2}{F} - \frac{2}{3 F^2}
- \frac{2}{3 F^5} \psi''' \left( 1+\frac{1}{F} \right)
 - \frac{1}{3 F^6} \psi'''' \left( 1+\frac{1}{F} \right)  \right](\alpha-1)+O \left( (\alpha-1)^2 \right)
\end{eqnarray}
whereas for large $\alpha$, we obtain 
the first corrections to the previous result (\ref{resdeltaof})
\begin{eqnarray}
\Delta_{0}(F,\alpha)
&& = 
 \frac{1}{F} + \frac{2}{F^4} \psi'' \left( 1+\frac{1}{F} \right)
+ \frac{1}{F^5} \psi''' \left( 1+\frac{1}{F} \right)
+ \frac{F^2}{\alpha^3}
+ \left[  3 F^2 - \frac{10}{3} F^3 + \frac{10}{3}
 \psi'' \left( 1+\frac{1}{F} \right) \right] \frac{1}{\alpha^4}
+O \left( \frac{1}{\alpha^5} \right)
\end{eqnarray}

For the case $t_w=0$, corresponding to $\alpha=\infty$ 
we have previously found a very simple relation (\ref{simpledeltafrommean})
between the two scaling functions describing the mean position
and the thermal width. Here, in the aging case $t_w>0$ with
finite $\alpha$, there doesn't seem to exist a simple generalization of 
(\ref{simpledeltafrommean}).

\section{ Various regimes for the response in the limit $\mu \to 0$  }

\label{reponseeffective}

In this Section, we translate our results
for the disorder averaged mean position (\ref{defxttwf},\ref{resxfa})
into the original unrescaled quantities, using the definitions
for $F= \beta f \xi(t,f)$ and for $\alpha(t,t_w,f)$ (\ref{defalpha}).
We have to distinguish various regimes, according to the relative
values of the times $(t,t_w)$ and of the characteristic
time scale $t_{\mu}(f)$ (\ref{tmuf}) associated to the force $f$.
 
\subsection{ Case $t_w  \ll   t_{\mu}(f)$ }

For $t_w  \ll   t_{\mu}(f)$, we obtain 
from (\ref{xfsmall},\ref{ashort}) 
the behavior in the sector $ t_w< t-t_w \ll   t_{\mu}(f) $ 
\begin{eqnarray} 
\overline{ <x> }(t,t_w,f) 
 \simeq \frac{\beta f }{2} (t-t_w)^{\frac{\mu}{1+\mu}} 
\left[ 1 - \left( \frac{ t_w }{ t- t_w} \right)^{\frac{2\mu}{1+\mu}} \right] 
\label{x1} 
\end{eqnarray}
 and from (\ref{xfbig},\ref{along}) the behavior
at long times  $ t-t_w \gg   t_{\mu}(f) $
\begin{eqnarray} 
\overline{ <x> }(t,t_w,f) \simeq \left( \frac{\beta f}{2}\right)^{\mu} (t-t_w)^{\mu}
\left( 1 - \frac{1}{2 } \left(  
\frac{ t_w^{\frac{1}{1+\mu}}}{ \frac{\beta f}{2}   (t-t_w) } \right)^{\mu}- \frac{1}{2 }  \left(   
\frac{ t_w^{\frac{1}{1+\mu}}}{ \frac{\beta f}{2}   (t-t_w) } \right)^{2\mu}\right)
\label{x2}
\end{eqnarray}

Moreover, at the beginning of the effective dynamics regime 
$\alpha \to 1$ (\ref{ashort}), 
the expressions (\ref{xfsmall}) and (\ref{xashort})
coincide and give
\begin{eqnarray} 
\overline{ <x> }(t,t_w,f)  \opsimeq_{t-t_w \to t_w} \beta f  t_w^{\frac{\mu}{1+\mu}} 
\left[  \left( \frac{ t- t_w}{ t_w } \right)^{\frac{\mu}{1+\mu}} -1 \right] 
\label{x3} 
\end{eqnarray}
whereas asymptotically when $\alpha \to \infty$ (\ref{along}),
the expressions (\ref{xfbig}) and (\ref{xalong}) give 
\begin{eqnarray}
\overline{ <x(t,f)> } \opsimeq_{t-t_w \gg \frac{2}{\beta f} t_w^{\frac{1}{1+\mu}}
} \left( \frac{\beta f}{2}\right)^{\mu} (t-t_w)^{\mu} 
 \label{x3bis} 
\end{eqnarray}

\subsection{ Case $t_w  \gg   t_{\mu}(f)$ }

For $t_w  \gg   t_{\mu}(f)$, the time sector $\alpha>1$ (\ref{along})
implies that $ t-t_w \gg   t_{\mu}(f) $ and
we obtain from (\ref{xfbig},\ref{along}) the behavior
at long times in time sector 
 $  t-t_w \gg  t_{\mu}(f)  $
and $ t-t_w > \frac{2}{\beta f} t_w^{\frac{1}{1+\mu}}   $
\begin{eqnarray} 
\overline{ <x> }(t,t_w,f)= \left( \frac{\beta f}{2}\right)^{\mu} (t-t_w)^{\mu}
\left( 1 - \frac{1}{2 } \left(   
\frac{ t_w^{\frac{1}{1+\mu}}}{ \frac{\beta f}{2}   (t-t_w) } \right)^{\mu}- \frac{1}{2 }  \left(  
\frac{ t_w^{\frac{1}{1+\mu}}}{ \frac{\beta f}{2}   (t-t_w) } \right)^{2\mu}\right)
\label{x4}
\end{eqnarray}

Here at the beginning of the aging regime 
$ t-t_w \to \frac{2}{\beta f} t_w^{\frac{1}{1+\mu}} $,
the expressions (\ref{xfbig}) and (\ref{xashort})
coincide and give
\begin{eqnarray} 
\overline{ <x> }(t,t_w,f)  \opsimeq_{t-t_w \to \frac{2}{\beta f} t_w^{\frac{1}{1+\mu}}}
 \frac{3}{2}  t_w^{\frac{\mu}{1+\mu}} 
\left[  \left(   
\frac{ \frac{\beta f}{2}   (t-t_w) }
{ t_w^{\frac{1}{1+\mu}}} \right)^{\mu} -1 \right] 
\label{x5} 
\end{eqnarray}
whereas at
at the end of the aging regime $\alpha \to \infty$ (\ref{along}),
the expressions (\ref{xfbig}) and (\ref{xalong}) again give (\ref{x3bis}).

\subsection{Discussion}

The results of this Section, that rely on the effective dynamics picture,
are valid at lowest order in the high disorder limit $\mu \to 0$,
and in the asymptotic aging regime where 
the two times are big $t \to \infty$, $t_w \to \infty$,
with the parameter $\alpha(t,t_w,f)>1$ (\ref{defalpha}) being fixed.
We have found
various interesting behaviors depending on the relative 
values of the parameters $(t,t_w,f)$.
Whenever they can be compared, our results agree
 with the scaling analysis presented in \cite{bertinjpreponse}.
In the next Section, we discuss the behavior of the response
in the time Sector $\alpha<1$, which 
is governed by rare events.

\section{ Response in the time sector $\alpha(t,t_w,f)<1$ from rare events  }

\label{rareevents}

As explained before, in the time sector $\alpha(t,t_w,f)<1$ (\ref{defalpha}),
 the effective dynamics governed by the decimation procedure
gives no contribution, and the response 
will thus be governed by rare events.
A similar situation was already found
in the RSRG studies 
 on the Sinai model \cite{us_sinai} and
and on the out-of-equilibrium dynamics
 of the random field Ising model \cite{us_rfim}.

\subsection{ Description of the rare events }

For the trap model considered here,
the `rare events' that are responsible for the response
in the time sector $\alpha(t,t_w,f)<1$,
can be describded as follows :
the particle, which is assumed to be trapped
in a renormalized trap $\tau>R(t_w,0)$ at time $t_w$
in the effective dynamics,
has actually a small probability 
to be at $t_w$ in a ``small" trap,  i.e. an already
decimated trap $\tau<R(t_w,0)$, for two reasons :

(i) when $\mu$ is small but finite, the particle can be found
at time $t_w$ in a trap $\tau<R_w$ with a probability of order $\mu$,
as explained in \cite{c_agingtrap},
where the corrections in $\mu$ with respect to the effective dynamics
of the unbiased trap model were studied in details.

(ii) when $t_w$ is large but not infinite, there is a small
probability that the particle is doing an excursion
(\ref{scaletout},\ref{scaletdiff}) at time $t_w$.

In both cases, the particle that happens to be in a ``small trap" at $t_w$
will respond to the external field
in the time sector $\alpha<1$.

\subsection{ Correction of order $\mu$ to the effective dynamics }

In the previous study on the unbiased trap model \cite{c_agingtrap},
we have studied in details the first corrections at order $\mu$ to
the effective dynamics. In particular, we have shown that
in the asymptotic time regime, the particle can be found with a probability
of order $\mu$ in the biggest trap $S$ contained in the interval $]M_-,M_+[$
between the two renormalized traps around the origin at scale $R(t_w,f=0)$.
In particular, we have obtained that the probability 
$\psi_{R_w}^{(\mu)}(\tau)$ to be at $t_w$
in a trap of trapping time $\tau$ was given in the domain $\tau<R_w$ 
at first order in $\mu$ by
 \begin{eqnarray}
\psi_{R_w}^{(\mu)}(\tau) \theta(\tau<R_w) \simeq 
\mu \frac{ R_w^{3/2} }{ \tau^{5/2}  }
K_1 \left( \sqrt{ \frac{R_w}{\tau} }  \right) 
K_2 \left( \sqrt{ \frac{R_w}{\tau} }  \right)
\label{psimu}
\end{eqnarray}
where we have dropped numerical factors of order $1$ (we refer the reader
to \cite{c_agingtrap} for precise results). 
In particular, this probability presents an essential singularity
at the origin
 \begin{eqnarray}
\psi_{R_w}^{(\mu)}(\tau)  \opsimeq_{\tau \to 0} 
\mu \frac{ R_w }{ \tau^2  } e^{-2  \sqrt{ \frac{R_w}{\tau} } }
\label{psimutau0}
 \end{eqnarray}
which means that it is very unlikely for the particle
to be trapped at $t_w$ in the biggest trap $S$ contained in the interval $]M_-,M_+[$ if the trapping time $\tau$ of $S$ is much smaller than $R_w$.

As a consequence, the integrated probability up to scale $R$ (with $R<R_w$)
 \begin{eqnarray}
 \int^{R} d\tau \psi_{R_w}^{\mu}(\tau)
\sim \mu \int_{\frac{1}{{\sqrt \alpha}}}^{+\infty} dz z^2 K_1(z) K_2(z) 
\end{eqnarray}
that is of order $\mu$ when $\alpha \sim 1$,
will also present an essential singularity
for small $\alpha=R/R_w$
 \begin{eqnarray}
 \int^{R} d\tau \psi_{R_w}^{\mu}(\tau)
\opsim_{\alpha \to 0} \mu \frac{1}{\sqrt \alpha} e^{ -\frac{2}{\sqrt \alpha}}
\label{integratedmu}
\end{eqnarray}

\subsection{ Probability to be doing an excursion at $t_w$ }

\subsubsection{ Probability to be doing an unsuccessfull excursion at $t_w$ }

From the discussion on unsuccessfull excursions,
the probability for a particle to be trapped in the vicinity of
a renormalized trap $\tau_0$ at scale $R$,
with neighbors at distances $l_{\pm}$
 reads when the external field vanishes (\ref{scaletout})
\begin{eqnarray}
P_{R}^{out}(\tau_0, l_+, \lambda_-) \simeq \frac{ t_{out} }{t_{in}}
\simeq \frac{ l_+ +l_- }{ 6 \tau_0 } 
\end{eqnarray}
After the average over $\tau_0$ with the measure (\ref{qrtau}) and over
the lengths (\ref{length}), we obtain
that the probability to be doing an unsuccessfull excursion
at time $t_w$ reads (\ref{defrw})
\begin{eqnarray}
P^{out}(t_w) \equiv \overline{ P_{R_w}^{out}(\tau_0, l_+, \lambda_-) } 
\sim  \mu R_w^{\mu-1} 
=  \mu t_w^{\frac{\mu-1}{1+\mu}}
\label{pouttw}
\end{eqnarray}
As expected, the probability of
these rare events is very small when $t_w$ is large since $\mu<1$.
In addition, there is a prefactor $\mu$ that makes this probability
even smaller in the limit $\mu \to 0$.

\subsubsection{ Probability to be doing a successfull excursion at $t_w$ }

The particle may be in a successfull excursion at $t_w$,
if it belongs to a renormalized trap that gets decimated
around the scale $R_w$.
From the discussion on successfull excursions,
the diffusion time for a length $l\sim R_w^{\mu}$
 reads when the external field vanishes
reads (\ref{diffusionlf},\ref{d0})
\begin{eqnarray}
t_{diff}(f=0,l) \simeq \frac{l^2}{6} \sim R_w^{2 \mu}   
\end{eqnarray}
This time window of width $(\Delta t)_w \sim R_w^{2 \mu}
 \sim t_w^{\frac{2 \mu}{1+\mu}}$ around $t_w$
 corresponds in RG scale to a window around $R_w \sim t_w^{\frac{1}{1+\mu}}$
of width $\Delta R_w \sim t_w^{- \frac{\mu}{1+\mu}} (\Delta t)_w
\sim t_w^{\frac{ \mu}{1+\mu}}$
As a consequence, the probability to be doing a successfull excursion at $t_w$
can be obtained from the probability to be in a trap $\tau \sim R_w$
(\ref{qrtau}) times the window width just estimated  
 \begin{eqnarray}
P^{diff}(t_w) \sim  q_{R_w}(R_w) \Delta R_w \sim \mu t_w^{\frac{ \mu-1}{1+\mu}}
\end{eqnarray}
The probability is again very small as expected, and happens to 
have exactly the same scale as (\ref{pouttw}).

\subsubsection{ Probability to be in a small trap $\tau$ during an excursion }

For a particle doing an excursion, we are now interested into the probability
$\psi^{exc}_{R_w}(\tau)$ to be in a trap of trapping time $\tau$.
We have obtained above that the total probability to be in a 
small trap $\tau<R_w$ behaves as 
 \begin{eqnarray}
\int^{R_w} d\tau \psi^{exc}_{R_w}(\tau) \sim   \mu t_w^{\frac{ \mu-1}{1+\mu}}
\sim \mu R_w^{\mu-1}
\end{eqnarray}
Assuming that the dependence of $\psi^{exc}_{R_W}(\tau)$ in $\tau$ is given by
$\tau q(\tau)$, i.e.
 the initial distribution (\ref{qtau}) weighed by the trapping time $\tau$
(i.e. the particle spends in each trap a time proportional to its trapping time), we obtain the estimation
 \begin{eqnarray}
 \psi^{exc}_{R_w}(\tau)    
\sim  R_w^{2(\mu-1)} \frac{ \mu }{ \tau^{\mu} } \theta(\tau<R_w)
\label{praretau}
\end{eqnarray}
which is rather flat for small $\mu \to 0$, in contrast with 
(\ref{psimutau0}). In particular,
the integrated probability up to scale $R$ (with $R<R_w$) reads
 \begin{eqnarray}
 \int^{R} d\tau \psi_{R_w}^{esc}(\tau)
\sim R_w^{2(\mu-1)} R^{1-\mu} =  R_w^{(\mu-1)} \alpha^{1-\mu}
\label{integratedesc}
\end{eqnarray}

\subsection{ Contribution to the response of these rare events }

When the initial condition
at $t_w$ is a `small' trap of trapping time $\tau$,
 the effective dynamics will become active again when 
$R(f,t-t_w)$ reaches $\tau$, i.e. when $\alpha$ reaches 
$\frac{\tau}{R_w}<1$, and the
corresponding contribution to the response reads 
 \begin{eqnarray}
(\overline{<x>} )_{\tau}(t,t_w,f) = 
\theta(R>\tau) R {\cal X}_{0} \left(F=\beta f R^{\mu} \right)
\end{eqnarray}
where the condition $\theta(R>\tau)$ means that the trap has been decimated
at scale $R$, and thus the response is given in terms of
the scaling function (\ref{calx0f}) found before for the case $t_w=0$.
Averaging over $\tau$ with the 
total probability $\psi_{R_w}(\tau)=\psi_{R_w}^{(\mu)}(\tau)+\psi_{R_w}^{exc}(\tau)$
coming from the two kinds of rare events describded above,
we obtain the leading term of the response in the sector $\alpha=R/R_w<1$ as
 \begin{eqnarray}
(\overline{<x>} )_{rare}(t,t_w,f)
= \int^{R} d\tau \psi_{R_w}(\tau)
(\overline{<x>} )_{\tau}(t,t_w,f) 
=  R {\cal X}_{0} \left(F=\beta f R^{\mu} \right) 
\left[ \int^{R} d\tau \psi_{R_w}^{(\mu)}(\tau)
+ \int^{R} d\tau \psi_{R_w}^{esc}(\tau)   \right]
\end{eqnarray}
Using the estimations of the integrated probabilities found
before (\ref{integratedmu},\ref{integratedesc}), 
we finally obtain that the contribution of excursions
dominate for small $\alpha<\alpha_w$, whereas the contribution
of the corrections in $\mu$ to the effective dynamics
are dominant for $\alpha_w<\alpha<1$. The scale of
the crossover value 
$\alpha_w$ can be estimated from 
the equality between (\ref{integratedmu}) and (\ref{integratedesc})
at leading order
\begin{eqnarray}
  \alpha_w \sim \frac{1}{ (\ln R_w )^2}
\end{eqnarray}

In conclusion, before 
the response of the effective dynamics in the sector $\alpha>1$ given in
(\ref{x1},\ref{x2},\ref{x4}), there exists a response in the sector
$\alpha<1$ as a consequence of rare events. 
For very small $\alpha<\alpha_w$, the response comes from
the particles doing excursions in anomalously small traps at $t_w$,
and it is reduced by a very small prefactor of order 
 $\mu t_w^{2(\mu-1)/(\mu+1)} (t-t_w)^{(1-\mu)/(\mu+1)}$
(\ref{integratedesc}). On the other hand, for $\alpha_w<\alpha<1$,
the response is governed by particles which are ``in delay"
with respect to the effective dynamics, as a consequence
of $\mu$ being finite, and the response is reduced by only
a factor of $\mu$.

\section{ Discussion of the Fluctuation-Dissipation Relation}

\label{fdtsection}

\subsection{ Linear response regime in a given sample
in the effective dynamics time sector}

From the diffusion front (\ref{ptwotimeinasample})
in a given sample characterized by $(\lambda^+,\lambda_-,\lambda^+,\lambda_-)$, we obtain at lowest order in $F$
the following results for the rescaled mean position 
\begin{eqnarray} 
\frac{ <x>(t,t_w,f) }{\xi(t-t_w,f) } = <Y>(\alpha,F) = \left[ \frac{\lambda^+ \lambda^-}{2}
 + \frac{ \lambda^+ \lambda^-_w + \lambda^+_w \lambda^-}{ 2 \alpha}  \right] F+
O(F^2)
\label{xinasample}
\end{eqnarray}
and the rescaled mean sqare displacement 
\begin{eqnarray} 
\frac{ <(x-x_w)^2>(t,t_w,f) }{\xi^2(t-t_w,f) } = <Y^2> (\alpha,F)= \left[ \lambda^+ \lambda^-
 + \frac{ \lambda^+ \lambda^-_w + \lambda^+_w \lambda^-}{  \alpha}  \right] +
O(F)
\label{y2inasample}
\end{eqnarray}
So the Fluctuation-Dissipation relation or Einstein relation
is valid in the whole time sector $\alpha>1$ as long as 
the linear response is valid and reads
in unrescaled quantities with $F=\beta f \xi(t-t_w,f)$,
\begin{eqnarray} 
 <x>(t,t_w,f) \opsimeq_{f \to 0} \frac{ \beta f}{2} <(x-x_w)^2>(t,t_w,0)
\label{fdtinasample}
\end{eqnarray}
This is in agreement with the scaling arguments and numerical simulations
presented in \cite{bertinjpreponse} and with the non-linear
Fluctuation Theorem discussed in \cite{c_nonlinearft},
that proves that the FDT relation is valid in any given sample
for arbitrary $\mu$.

The validity of the FDT relation for the trap model
in its aging phase is nevertheless quite remarkable, 
 since the  
dynamics is completely out-of-equilibrium :
indeed, in the effective dynamics,
the weights of the two important traps are 
not given by Boltzmann
factors, they don't even depend on the energies of these two traps,
but they are given by the probabilities
to reach one before the other one, and they thus
only depend on the distances to the origin!
This example, with the explicit expressions in a given sample
(\ref{xinasample},\ref{y2inasample}), thus shows that
the validity of the FDT relation in the linear response regime
does not imply that the system is at equilibrium or even near equilibrium.

\subsection{ Non-linear response  
in the asymptotic aging sector $\alpha \to \infty$ }

In the asymptotic aging sector $\alpha \to \infty$,
which also corresponds to the case 
where the external field is applied from the very begining $t_w=0$,
we have found a very simple relation (\ref{simpledeltafrommean})
between the scaling functions for the mean displacement
and the thermal width which is valid for arbitrary $F$
and in particular in the whole non-linear response regime.
However, this relation found for disorder averaged quantities
does not seem to have a simple interpretation, since 
in a given sample, there is not such a relation 
between the rescaled mean position 
and the rescaled thermal width
that are given by  (\ref{pMM})
\begin{eqnarray}
\frac{ <x(t,f)> }{ \xi(t,f) } && = 
 \frac{ \lambda^+ (1-e^{-F \lambda^-}) - \lambda^- e^{-F \lambda^-} (1-e^{-F  \lambda^+} ) }
{ 1- e^{-F ( \lambda^++\lambda^-) }} \\
\frac{ < \Delta x^2(t,f)> }{ \xi^2(t,f) }  &&  = 
 ( \lambda^+ + \lambda^-)^2
\frac{ e^{- F \lambda^-} (1-e^{-F  \lambda^+} ) ( 1-e^{-F \lambda^-} ) }
{ ( 1- e^{-F ( \lambda^++\lambda^-) } )^2}
 \end{eqnarray}
Moreover, we have not found an equivalent
relation when $\alpha$ is finite (\ref{thermalwidthalpha}).
Nevertheless, after the average over the samples, we
have obtained the following simple property at very long times 
in the aging regime $\alpha \to \infty$ (\ref{along}),
i.e. in the non-linear response regime (\ref{moyxfbig},\ref{deltax2tlarge})
\begin{eqnarray}
\lim_{t \to \infty} \left( \frac{ \overline{ <x> (t,t_w,f)} }
{ \overline{ <\Delta x^2(t,t_w,f)> } }  \right) = \beta f 
 \end{eqnarray}
whereas, for comparison, there is a factor $(1/2)$ in the FDT 
relation of the linear response regime (\ref{fdtinasample})
\begin{eqnarray}
\left( \frac{ \overline{ <x> (t,t_w,f)} }
{ \overline{ <\Delta x^2(t,t_w,f)> } }  \right) \opsimeq_{f \to 0}
 \frac{\beta f}{2}
   \end{eqnarray}

As a comparison, in the pure trap model, it is immediate to obtain
the equations for the mean displacement 
$( d<n>_{pure}/dt = 2 \sinh \frac{\beta f}{2} $
and for the thermal width
$ d< \Delta n^2>_{pure}/dt = 2 \cosh \frac{\beta f}{2} $
so that their ratio is simply
\begin{eqnarray}
\frac{ <n>_{pure} (t)}{ < \Delta n^2>_{pure} (t)} 
=  \tanh \frac{\beta f}{2} \opsimeq 
\frac{\beta f}{2}
 \end{eqnarray}
for arbitrary time,
in the regime of small asymetry we are interested in (\ref{hof}).

\section{Conclusion}

\label{conclusion}

We have studied in details the dynamics
 of the one dimensional disordered trap model 
when an external force is applied from the
very beginning at $t=0$, or only after a waiting time $t_w$,
in the linear as well as in the non-linear response regime.
Using a disorder-dependent real-space renormalization procedure 
that becomes exact in the limit of strong disorder $\mu \to 0$,
we have shown that 
the diffusion front in each sample consists in two delta peaks, which are completely out of equilibrium with each other, since their weights
represent the probabilities to reach one before the other one.
 The statistics of the positions and weights of these delta peaks over the samples was then used to obtain explicit results for many observables,
such as the diffusion front, the mean position,
the thermal width, the localization parameters and
 the two-particle correlation function.
Since the renormalization procedure is defined 
sample by sample, our approach provides a very clear insight
into the important dynamical processes.

 From a more general perspective, it seems
that up to now, the studies on the response of aging systems to an external field
have been mainly restricted to
 the linear response regime \cite{reviewleticia,reviewritort}, which holds
for fixed times $(t_w,t)$ in the limit of vanishing field $f \to 0$.
However, as in the trap model discussed here, it 
should be expected in a broad
class of aging systems that, for a fixed small field $f$, 
the validity of the linear response regime
is limited in the time sector for $(t_w,t)$
by a characteristic time $\tau(f)$
depending on the external field. Indeed, it seems rather natural
that an external field, even if it is arbitrarily small, will, 
for sufficiently long times, drive the system into
a configurational landscape which is completely different from the initial one.
So in the asymptotic time regime
beyond the characteristic scale $\tau(f)$, 
 the response will always be governed by non-linear effects. 
For the special case $t_w=0$,
the full response including these non-linear effects
 has already been studied for the Sinai model \cite{us_sinai}
as well as in the coarsening
dynamics of the random field Ising model \cite{us_rfim},
via the RSRG approach : in both cases, as in the trap model,
the field introduces a characteristic time separating 
the linear response regime from a non-trivial aging regime
with non-linear effects. This scenario should more 
generally apply to other coarsening dynamics. 
However, in numerical studies on domain growth processes,
to get better results for the linear response regime
at large times,
it is usually the response to a random field that is measured
\cite{barrat,corberi}, and not the response to a constant field,
that would favor one of the phase and induces non-linear effects
rapidly. Since this choice of a random field for
domain growth processes, is in some sence the equivalent
of a constant field for spin-glasses \cite{barrat},
it seems that the non-linear response
of spin-glasses will be very different from coarsening systems.
For instance, in the dynamics of the
spherical Sherrington-Kirkpatrick spin-glass model \cite{leticiadavid},
the magnetic field introduces a characteristic time
that separates the aging dynamics of the linear
response regime from an equilibrium dynamics at large times :
here, the magnefic field does not lead at large times to a non-trivial
aging regime with non-linear effects in the field,
but rather gives rise to an interrupted aging phenomenon.
However, this scenario is not expected to hold for other spin-glasses
such as the usual SK model \cite{leticiadavid}, 
if one considers the number of metastables states
in a field \cite{dean}.   
In conclusion, the understanding of the non-linear effects 
thar arise at large times in the response of aging systems
is still very incomplete and should give rise to further studies
in the future.

 \begin{acknowledgments}

It is a pleasure to thank E. Bertin, J.P. Bouchaud, G. Biroli
and J. M. Luck for useful discussions.

\end{acknowledgments}

\appendix

\section{ Statistical properties of excursions in the presence of a field }

\label{appexcursions}

As explained in the text, to study the excursions in the 
renormalized landscape in the presence of a field, we have to study the following standard problem :
what is the probability distribution
$P_{t}(x)$ of the time $t$ of the first-passage at $x=0$
without having touched the other boundary $x=l$ before,
for a pure random walk starting at $x$ with
 the asymetry $h=h(f)>0$ (\ref{q+-f}) ?

For $x=1,...,l-1$, the probability distribution
$P_{t}(x)$ satisfies the equation
\begin{eqnarray}
\partial_t P_{t}(x) =  (1+h) P_{x-1}(t) + (1-h) P_{x+1}(t) - 2 P_{t}(x) 
\end{eqnarray}
with the boundary conditions $P_0(t)= \delta(t)$ and $P_l(t)=0$.
So the Laplace transform with respect to $t$
\begin{eqnarray}
{\hat P}_x(s) \equiv \int_0^{+\infty} dt e^{-s  t }  P_{t}(x)
\end{eqnarray}
satisfies
\begin{eqnarray}
(1+h){\hat P}_{x+1}(s) + (1-h) {\hat  P}_{x-1}(s) 
- (2+s) {\hat  P}_x(s) = 0
\end{eqnarray}
for $x=1,...,l-1$
with the boundary conditions 
${ \hat P}_0(s)= 1$ and ${\hat P}_l(s)=0$.

The solution reads
\begin{eqnarray}
 {\hat  P}_x(s) && =  \frac{ \rho_+^l(s) \rho_-^x(s) - \rho_-^l(s) \rho_+^x(s) }
{\rho_+^l(s)- \rho_-^l(s) } 
\label{pxs}
\end{eqnarray}
in terms of the roots
\begin{eqnarray}
\rho_{\pm}(s) = \frac{2+s \pm \sqrt{s^2+4s+4 h^2} }{2 (1+h)}
\end{eqnarray}
The series expansion in $s$ then yields the first moments
\begin{eqnarray}
\theta_k(x) \equiv \int_0^{+\infty} dt t^k P_{t}(x)
\label{moments}
 \end{eqnarray}

\subsection{ Escape probabilities }

For $s=0$, the roots become (we assume $h>0$) in terms of the bias $f$
(\ref{q+-f})  
\begin{eqnarray}
\rho_{+}(0) && = 1 \\
\rho_{-}(0) && = \frac{1-  h }{1+  h } = e^{- \beta f}
\end{eqnarray}
and the probability to reach $0$ before $l$ when starting at $x$
thus reads
\begin{eqnarray}
\theta_0(x) =  {\hat  P}_x(s=0)  = 
 \frac{  e^{- \beta f x } - e^{- \beta f l }  }
{1- e^{- \beta f l } }
\end{eqnarray}

\subsubsection{ Escape probability along the drift}

When starting at $x=1$, the probability to reach $x=l$ 
without any visit to $x=0$ reads
\begin{eqnarray}
p_e^+(l,f)=1-\theta_0(x=1) = 
 \frac{ 1 - e^{- \beta f  }   }
{1- e^{- \beta f l } } 
\label{escape+}
\end{eqnarray}
Here we are interested in the regime $\beta f \ll 1$ and $l \gg 1$, where
the escape probability takes the scaling form
\begin{eqnarray}
p_e^+(l,f) \opsimeq  \frac{1}{l} E^+ ( u = \beta f l )
\label{scalingescape+}
\end{eqnarray}
with the scaling function
\begin{eqnarray}
E^+(u) = \frac{ u}{1-e^{-u}}
\end{eqnarray}
In particular, $E^+(u \to 0) \to 1$
corresponds to the unbiased case
 where the escape probability is simply $1/l$.
In the other limit where $u = \to +\infty$, we have $E^+(u) \simeq u $
and the escape probability becomes $ (\beta f) $
i.e. it is proportional to the drift and independent of $l$.

\subsubsection{ Escape probability against the drift}

When starting at $x=l-1$, the probability to escape to $0$
without without any visit to $x=l$ reads 
\begin{eqnarray}
p_e^-(l,f)=\theta_0(x=l-1) =  
 \frac{ ( 1 - e^{- \beta f  } ) e^{- \beta f (l-1) }  }
{1- e^{- \beta f l } } 
\label{escape-}
\end{eqnarray}
which varries as it should between $p_e^-(l,f \to 0)=1/l$ for the unbiased case
and $p_e^-(l,f \to 1)=0$ for the directed case.
In the regime $\beta f \ll 1$ and $l \gg 1$, 
the escape probability takes the scaling form
\begin{eqnarray}
p_e^-(l,f) \opsimeq  \frac{1}{l} E^- ( u= \beta f l  )
\label{scalingescape-}
\end{eqnarray}
with the scaling function
\begin{eqnarray}
E^-(u) = \frac{ u}{e^{u}-1}
\end{eqnarray}
In particular, $E^-(u\to 0) \to 1$
corresponds to the unbiased case
 where the escape probability is simply $1/l$.
In the other limit where $u  \to +\infty$, we have $E^-(u) \simeq u
e^{- u} $
and the escape probability becomes exponentially small
$p_e^-(l,f) \opsimeq  \beta f e^{- \beta f l}$ .

\subsection{ Mean time for unsuccessfull excursions }

The expansion at first order in $s$ of (\ref{pxs})
yields (\ref{moments}) the mean time to reach $0$
without any visit to $x=l$ when starting at $x$
\begin{eqnarray}
\theta_1(x) =  \frac{ 1+e^{- \beta f  } }{ 1- e^{- \beta f  } } 
\left[
 \frac{ x
( e^{- \beta f x } + e^{- \beta f l } )  }
{ 2  ( 1- e^{- \beta f l } ) }
- \frac{ l e^{- \beta f l } 
( 1- e^{- \beta f x }  )  }
{   ( 1- e^{- \beta f l } )^2 } \right]
\end{eqnarray}

\subsubsection{ Unsuccessful Excursion along the drift}

For $x=1$, in the limit $\beta f\ll1$ and $l\gg1$,
the mean time of unsuccessful excursions along the drift 
takes the scaling form  
\begin{eqnarray}
\theta_{us}^+(f,l) = \theta_1(x=1) \simeq l \Theta(u=\beta fl)
\label{meanplus}
\end{eqnarray}
where the scaling function
\begin{eqnarray}
 \Theta(u) = \frac{ 1 - 2u e^{-u} -e^{-2 u} }{  u (1-e^{- u})^2}
\label{thetaexcursion}
\end{eqnarray}
interpolates between
\begin{eqnarray}
 \Theta(u) =\frac{1}{3} - \frac{u^2}{90}  +O(u^4) 
\label{theta0}
\end{eqnarray}
for the unbiased case, where the mean time is $(l/3)$, and
\begin{eqnarray}
 \Theta(u) \opsimeq_{u \to \infty} \frac{1}{u} 
\label{thetainfty}
\end{eqnarray}
 where the mean time is $1/(\beta f)$.

\subsubsection{ Unsuccessful Excursion against the drift}

The unsuccessful excursions against the drift
have the same properties 
\begin{eqnarray}
\theta_{us}^-(f,l) = \theta_{us}^+(f,l)
\label{samepm}
\end{eqnarray}

\subsection{ Mean time for the successful excursions }

Similarly, we find that the mean time needed to reach $x=l$
when starting at $x=1$ for a random walk conditioned not to visit $x=0$
takes the scaling form
\begin{eqnarray}
t_{diff}^+(f,l) \simeq l^2 D( u=\beta fl) ) 
\label{diffusionlf}
\end{eqnarray}
where the scaling function
\begin{eqnarray}
 D (u) = 
\frac{ u-2 +(u+2)  e^{-u} }{  u^2 (1-e^{-u})} 
\label{functionDu}
\end{eqnarray}
interpolates between
\begin{eqnarray}
 D (u) = 
\frac{1}{6} +O(u^2) 
\label{d0}
\end{eqnarray}
for the unbiased case, where the mean diffusion time is $(l^2/6)$, and
\begin{eqnarray}
  D (u)
= \frac{1}{  u} +O\left( \frac{1}{u^2} \right)
\label{dinfty}
\end{eqnarray}
where the mean diffusion time is $l/(\beta f)$.

Similarly, we find that 
the mean time for a successful excursion against the drift 
has the same properties

\begin{eqnarray}
t_{diff}^-(f,l) = t_{diff}^+(f,l)  
\label{diff-+}
\end{eqnarray}

\end{document}